\documentclass[useAMS,usenatbib]{mn2e}
\pdfoutput=1
\usepackage{graphicx}
\usepackage{subfigure}
\usepackage[english]{babel}
\usepackage{amssymb}
\usepackage{xspace}
\usepackage{amsmath}
\usepackage[draft]{hyperref}
\usepackage{color}
\usepackage{pdflscape}
\usepackage{afterpage}

\def\hii{\ifmmode {\mbox H{\scshape ii}}\else H{\scshape ii}\fi\xspace}
\def\Galex{{\it GALEX}\xspace}
\def\IRXB{IRX -- $\beta$\xspace}
\def\Angstrom{\AA\xspace}

\title [Dissecting the \IRXB dust attenuation relation]{Dissecting the
  \IRXB dust attenuation relation: exploring the physical
  origin of observed variations in galaxies}
\author[G. Popping, A. Puglisi, and C. A. Norman]{Gerg\"o
  Popping$^{1}$\thanks{E-mail:
    gpopping@eso.org}, Annagrazia Puglisi$^{1, 2}$, and Colin A. Norman$^{3,4}$\\
$^{1}$European Southern Observatory, Karl-Schwarzschild-Strasse 2,
85748, Garching, Germany\\
$^{2}$Dipartimento di Fisica e Astronomia, Università di Padova,
vicolo dell`Osservatorio 2, I-35122 Padova, Italy\\
$^{3}$Department of Physics and Astronomy, Johns Hopkins University,
Baltimore, MD 21218, USA\\
$^{4}$Space Telescope Science Institute, 3700 San Martin Drive, Baltimore, MD 21218, USA\\}
\begin{document}

\maketitle

\begin{abstract}
The use of ultraviolet (UV) emission as a tracer of galaxy
star-formation rate (SFR) is hampered by dust obscuration. The
empirical relationship between UV slope, $\beta$, and the ratio
between far-infrared and UV luminosity, IRX, is commonly employed to
account for obscured UV emission. We present a simple model that
explores the physical origin of variations in the \IRXB dust
attenuation relation. A relative increase in FUV compared
to NUV attenuation and an increasing stellar population age cause variations
towards red UV slopes for a fixed IRX. Dust geometry effects
(turbulence, dust screen with holes, mixing of stars within the dust
screen, two-component dust model) cause variations towards blue UV
slopes. Poor photometric sampling of the UV spectrum causes
additional observational variations. We provide an analytic approximation for the \IRXB relation
invoking a subset of the explored physical processes (dust type,
stellar population age, turbulence). We discuss observed variations in
the \IRXB relation for local (sub-galactic scales) and high-redshift
(normal and dusty star-forming galaxies, galaxies during the epoch of
reionization) galaxies in the context of the physical processes explored in our
model. High spatial resolution imaging of the UV and sub-mm
emission of galaxies can constrain the \IRXB dust
attenuation relation for different galaxy types at different
epochs, where different processes causing variations may dominate. These constraints will allow the use of the \IRXB relation to estimate intrinsic SFRs of galaxies, despite the lack of a universal relation.

\end{abstract}


\section{Introduction}
A reliable measurement of the star-formation rate (SFR) of galaxies is
a key component of our understanding of galaxy formation and
evolution.  A commonly used approach to measure the SFR of galaxies accounts
for the ultraviolet emission from short-lived massive stars \citep{Kennicutt2012}. This approach is
especially popular for high-redshift galaxies up to redshifts of
$z\sim 10$, where broad-band UV
photometry is available through deep field studies
\citep[e.g.,][]{Bouwens2011,Dunlop2012,Ellis2013,Oesch2014,Oesch2015,Bouwens2015,Mcleod2015}. The
use of UV as a SFR tracer is hampered by the absorption of UV radiation by dust in
the interstellar medium (ISM). The absorbed radiation is re-emitted in
the infrared (IR), and ideally far-IR measurements complementary to UV detections are available to account for the absorbed
emission. Unfortunately,  especially at $z>3$ IR information is not always available due to
various reasons (e.g., detection limits and source confusion in
\emph{Spitzer Space Telescope} and
\emph{Herschel Space Observatory} bands) and
alternative approaches to account for the absorbed
UV emission have to be used.

A commonly used approach to account for the absorbed UV
emission is the empirical relation between the rest-frame UV continuum slope
$\beta$ (where $F_\lambda \propto \lambda^\beta$) and the ratio
between the infrared and UV luminosity (IRX) in local (blue starburst) galaxies 
\citep{Calzetti1997,Meurer1999, Overzier2011, Takeuchi2012}. This
relation correlates the UV slope to dust extinction, which then is used
to correct the UV photometry for dust obscuration and derive the
intrinsic (dust-obscured plus unobscured) SFR of galaxies.  If the
scatter in this relation is minimal over a range of galaxy types over
cosmic time, it
can serve as a straightforward way to infer the
intrinsic SFR of galaxies when only UV measurements are available. The
\IRXB relation has been widely used to estimate the intrinsic SFR in high-redshift galaxies
for which IR data is not available
\citep[e.g.,][]{Bouwens2011,Ellis2013,Oesch2014,Oesch2015,Bouwens2015,Mcleod2015}. Accounting
for dust obscured star formation at high redshifts is essential, as observations and models suggest that galaxies
at $z>7$ can already have significant reservoirs of dust
\citep[e.g.,][]{Watson2015, Popping2016, Laporte2017, Strandet2017}.
The use of the \IRXB relation for high-redshift galaxies is supported
by observations that showed that the Meurer et al. \IRXB relation is
broadly appropriate for samples of main-sequence galaxies at $z\sim 2 - 4$
\citep{Reddy2008,Pannella2009,Reddy2010,
  Reddy2012,Heinis2013,To2014,Coppin2015,Alvarez-Marquez2016,Puglisi2016,Bourne2017, Fudamoto2017}.

\begin{figure*}
\includegraphics[width = 1.0\hsize]{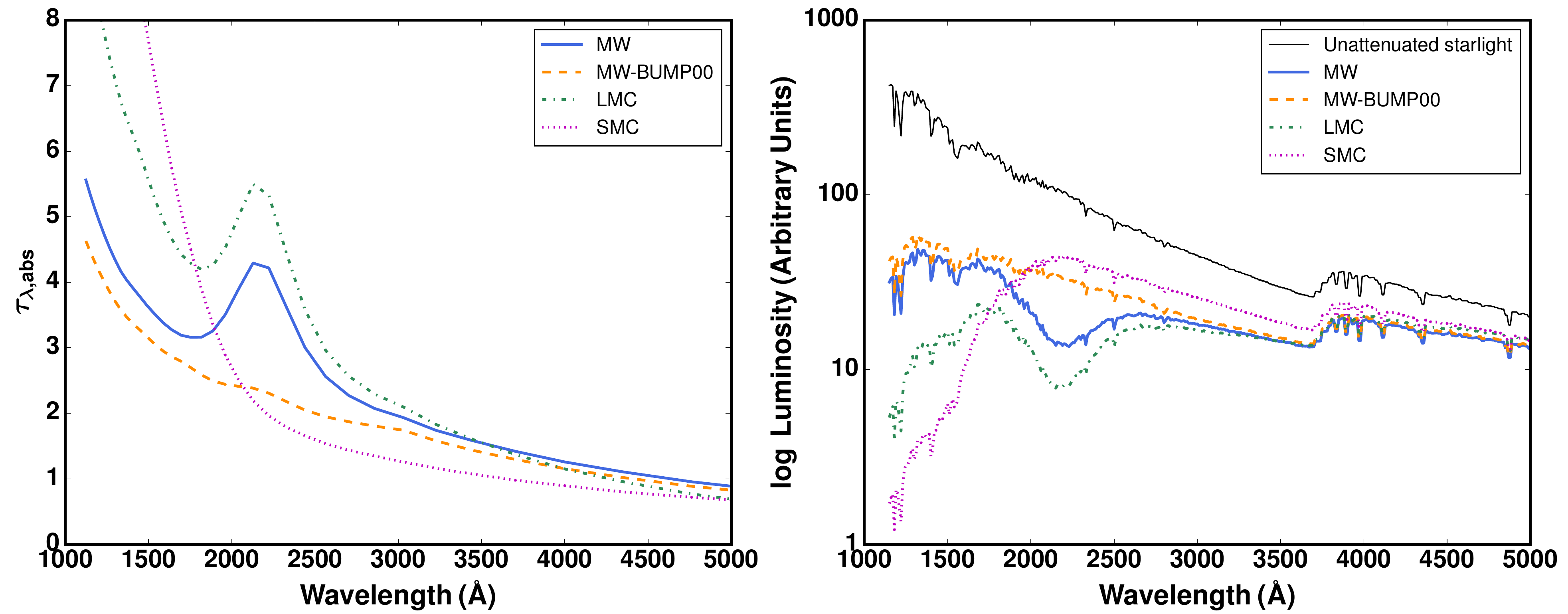}
\caption{Left: The dust extinction curves applied in this work for a
  mean homogeneous optical depth in the V band of one ($<\tau_{\text
    V}> = 1$) and assuming a uniform medium ($\mathcal{M} =0$) and
  $R_{\rm s}/R_{\rm d} =0$. Right: The unobscured (black) and obscured
  (coloured) stellar emission spectra of a single
  burst population with an age of 10 Myr and solar metallicity. The
  obscured spectra are based on the attenuation curves shown in the
  left panel.\label{fig:attenuation_law}}
\end{figure*}

Despite its promising nature,  local and
high-redshift studies have showed that some galaxies deviate from the \IRXB relation. Nearby ultra-luminous infrared galaxies (ULIRGS) have large
values of IRX with a low value of $\beta$, corresponding to a blue UV
continuum slope \citep{Goldader2002}. Observations of ordinary spiral
galaxies in the local Universe showed larger values for $\beta$ (corresponding to a redder UV continuum slope) for a given
IRX compared to the \citet{Meurer1999} relation
\citep{Kong2004,Buat2005,Grasha2013}. The latter was explained by older and less
massive stars contributing to the near-UV emission of galaxies. Metal
poor systems such as the Magellanic clouds have redder colours than
metal-rich systems \citep{Bell2002,Bell2002B}. There are many other
works that have
explored variations in the \IRXB relation and their physical cause in
local galaxies \citep[e.g.,][]{Burgarella2005,Seibert2005, Cortese2006,
  GildePaz2007, Johnson2007, Panuzzo2007, Cortese2008, Siana2008,Boquien2009,
  Munoz-Mateos2009,Wijesinghe2011, Boquien2012,Ye2016}. High-redshift
observations have demonstrated that dusty star-forming galaxies (DSFGs) have bluer UV
continuum slopes for a given IRX
\citep[increasingly deviating from the Meurer et al. relation as a
function of IR luminosity, e.g.,][]{Oteo2013,Casey2014,Bourne2017}. High-redshift Lyman
Break Galaxies are observed to have redder UV continuum slopes $\beta$
for a given IRX value, more consistent with a Small Magellanic Cloud
(SMC) or Large Magellanic Cloud (LMC) attenuation curve than the
canonical \citet{Meurer1999} relation
\citep[e.g.,][]{Capak2015,Bouwens2016,Koprowski2016,Pope2017, Smit2017}. A
similar conclusion was reached for normal star-forming galaxies at
$z\sim 2$ \citep{Reddy2012,Reddy2017} and for lensed Lyman Break
Galaxies at $z\sim3$ \citep{Siana2008,Siana2009}. Recently, \citet{Fudamoto2017}
found that IRX decreases as a function of
look-back time for normal star-forming galaxies at $z\sim4-6$ at a given UV slope $\beta$.

A reliable use of the \IRXB relation to estimate the dust-obscured
star formation (SF) in galaxies requires a detailed understanding of
the physical origin of the aforementioned variations. Multiple
theoretical efforts have addressed the origin of observed variations in the \IRXB
relation \citep[e.g.,][]{Granato2000,Ferrara2016,Mancini2016,Cullen2017, Narayanan2017,Safarzadeh2017}. These works have highlighted stellar population age,
variations in the dust attenuation curve, the temperature of the dust,
and the geometry of the dust distribution as possible origins for observed deviations from the
Meurer et al. \IRXB relation.  

In this work we present a simple model to systematically
explore how different properties of the stellar population (age), different
properties of the absorbing dust screen (dust attenuation curve, turbulence, geometry, mixing
of stars and dust, and a two-component dust model), and observational effects
affect the \IRXB relation of galaxies. We present our methodology in
Section \ref{sec:model} and a detailed analysis of the physical origin
of variations in the \IRXB relation in Section \ref{sec:analysis}. We discuss our
results in Section \ref{sec:discussion}, focusing on the physical
origin of the \IRXB relation, observed variations in the \IRXB
relation in the context of our model results, an analytic
approximation of the \IRXB relation, and a future outlook towards
reliably using the \IRXB relation as a tracer of dust-obscured
star-formation. We summarise our results in Section \ref{sec:summary}.

\section{Model description}
\label{sec:model}
In this work we use a simple model in which we place a screen of dust
in front of a single burst simple stellar population. Throughout the
paper we vary the properties of the screen and the properties of the
stellar population.

\subsection{Stellar population model}
We use the \texttt{Starburst99} \citep{Leitherer1999} simple stellar
population model to create stellar spectra. The Spectral Energy
Distributions (SEDs) are calculated as simple stellar population models
with a fixed stellar mass at different ages and metallicities. The
model assumes a \citet{Kroupa2002}
initial mass function and the Geneva Isochrones
without rotation \citep{Ekstrom2012,Georgy2013}. Unless noted
differently we use an SED generated assuming a stellar age of 10 Myr
and solar metallicity.

\subsection{Dust screen}
The attenuation by dust is modelled using the attenuation curves from
\citet{Seon2016}. Seon \& Draine apply radiative transfer models in a
spherical, clumpy interstellar medium to investigate the attenuation
of starlight. In this model the turbulent dusty medium and stellar
sources are spherically distributed within a radius of $R_{\rm d}$ and
$R_{\rm s}$ for dust and stars,
respectively. Photon sources are uniformly distributed within the
stellar sphere. The radial extent of the stellar sources is varied
within the range $0<R_{\rm s}/R_{\rm d} < 1$. $R_{\rm s}/R_{\rm
  d}  = 1$ corresponds to a case where the photon sources are
uniformly distributed over the dusty sphere (photon sources are
uniformly mixed), whereas $R_{\rm s}/R_{\rm
  d} =0$ corresponds to a medium where the photon sources are surrounded by
a dusty cloud \citep{Seon2016}. The latter case can be seen as a
screen of dust located between the photon sources and the
observer. We point the reader to Figure 4 in
  \citet{Seon2016} for a schematic representation of the spatial
  distribution between dust and photon sources.

The Seon \& Draine model provides attenuation optical depths due to
absorption and scattering from approximately 1000 \Angstrom to 3 $\mu$m, as a
function of mean homogeneous V-band optical depth ($<\tau_{\text v}>$; from
$<\tau_{\text v}> = 0.1$ to $<\tau_{\text v}>=20$). \citet{Seon2016}
define the homogeneous
 V-band optical depth as the centre-to-edge optical depth
 of a cloud with a constant density and radius $R_{\rm d}$. Photon packages
that escape the spherical dust cloud are recorded as $f^{\rm esc}_\lambda$ as a
function of wavelength. Depending on the optical depth of
the system,  between $10^5$ and $10^7$
photon packages are used for each wavelength. The attenuation optical
depth is then
calculated as $\tau^{\rm att}_\lambda = -\ln{(f^{\rm
    esc}_\lambda)}$. The attenuation optical depth in the V-band is often less than
the homogeneous V-band optical depth, as a significant fraction of the
scattered light does escape the dust sphere towards the
observer. In this work we adopt attenuation curves based on the
following dust types from \citet{Weingartner2001}: Milky Way (MW), Milky Way without the
2175 \Angstrom bump (MW-BUMP00), Large Magellanic Cloud (LMC), and
Small Magellanic Cloud (SMC).

The \citet{Seon2016} model takes the turbulence of the
screen of dust into account. When no turbulence is present the screen
has a uniform distribution of dust and the optical depth seen by the
stellar light is identical at any point on the screen surface. Under the influence of turbulence the
mean optical depth  in the V-band $<\tau_{\text v}>$ of a screen of
dust stays the same, but locally the optical depth increases or decreases \citep[see
also for example][]{Fischera2003, Fischera2005}. This naturally adds geometry to the
screen of dust and changes the attenuation properties. The \citet{Seon2016} model
explores the level of turbulence of the dust screen from a uniform
screen (Mach number $\mathcal{M} = 0$) to a highly turbulent screen (Mach
number $\mathcal{M}=20$). We interpolate between wavelength $\lambda$, mean V-band optical
depth, $R_{\rm d}/R_{\rm s}$, and Mach number to derive optical depths as a function of the
aforementioned properties. 

The attenuated stellar emission at any
wavelength, mean V-band optical depth, Mach number, and $R_{\rm d}/R_{\rm s}$ is given as
\begin{multline}
I_{*, \rm{att}}(\lambda, <\tau_{\text v}>, \mathcal{M}, R_{\rm d}/R_{\rm s}) = \\I_*(\lambda)\,\times\,
e^{-\tau^{\rm att}(\lambda,<\tau_{\text v}>, \mathcal{M},R_{\rm d}/R_{\rm s})},
\end{multline}
where  $I_*(\lambda)$ is the unattenuated stellar
emission. 

As an example we show the attenuation
curve for the MW, MW-BUMP00, LMC, and SMC in the left panel of Figure
\ref{fig:attenuation_law} (assuming $<\tau_{\text v}> = 1$, $\mathcal{M}=0$, and
$R_{\rm d}/R_{\rm s} =0$). There are clear differences between the
attenuation curves, especially in the level of attenuation at UV
wavelengths (up to $\sim$ 3000 \Angstrom). Besides the presence of the
2175 \Angstrom bump in the MW and LMC curves, the LMC and SMC curves
show much more attenuation  in the FUV regime. The LMC curve shows an
elevated level of attenuation compared with the MW curves at
wavelengths $\lambda < 4500$\Angstrom, whereas the SMC curve has an
elevated level of attenuation at wavelengths $\lambda <
2000$\Angstrom. The right panel of Figure \ref{fig:attenuation_law}
shows the unattenuated stellar emission of a single burst population
with solar metallicity and an age of 10 Myr, and the attenuated
spectra for the MW, MW-BUMP00, LMC, and SMC attenuation curves
presented in the left panel of Figure \ref{fig:attenuation_law}. The 2175 \Angstrom feature is
clearly visible in the attenuated spectra for the MW and LMC
attenuation curves. The strong increase in absorption for the SMC
attenuation curve at wavelengths less than 2000 \Angstrom reflects
itself by a steeply rising attenuated stellar emission in the same
wavelength range. 

Unless noted differently, we use a uniform dust
screen ($\mathcal{M}=0$) with $R_{\rm d}/R_{\rm s} = 0$ in the remaining of the
paper. In this work we always assume a single burst stellar population,
rather than the more realistic composite stellar
  populations galaxies consist of.

\begin{figure*}
\includegraphics[width = 0.65\hsize]{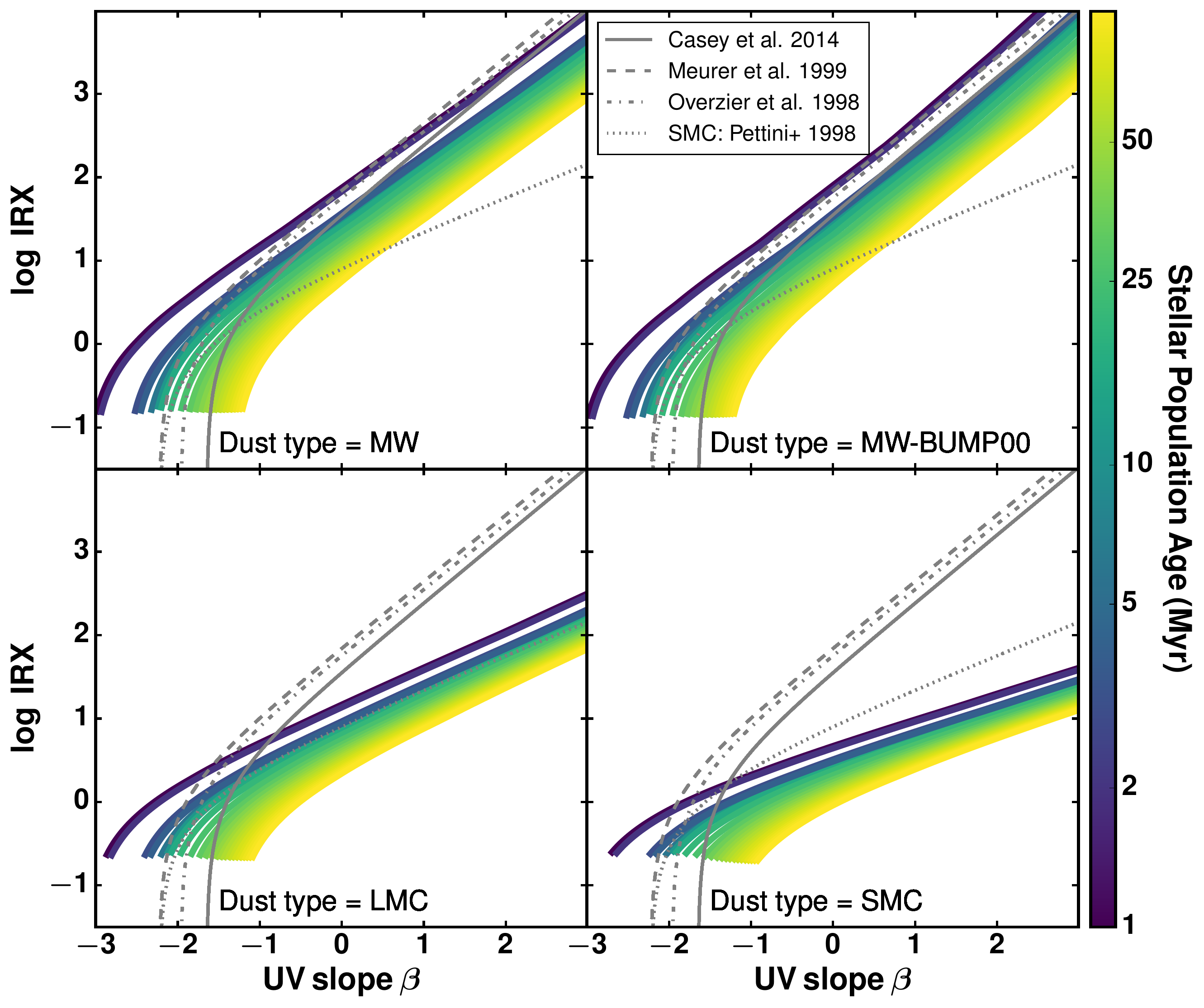}
\caption{The \IRXB relation for different dust types and a
  varying age of the stellar population compared to
  observed \IRXB relations \citep{Meurer1999,Overzier2011,Casey2014, Pettini1998}. Larger
  values of IRX correspond to an increased optical depth. Older populations shift the relation to larger
  values of UV-slope $\beta$. An increased FUV attenuation component
  (LMC and SMC dust types) flattens the \IRXB relation.
\label{fig:fiducial_beta_IRX}}
\end{figure*}
\begin{figure}
\centering
\includegraphics[width = 0.9\hsize]{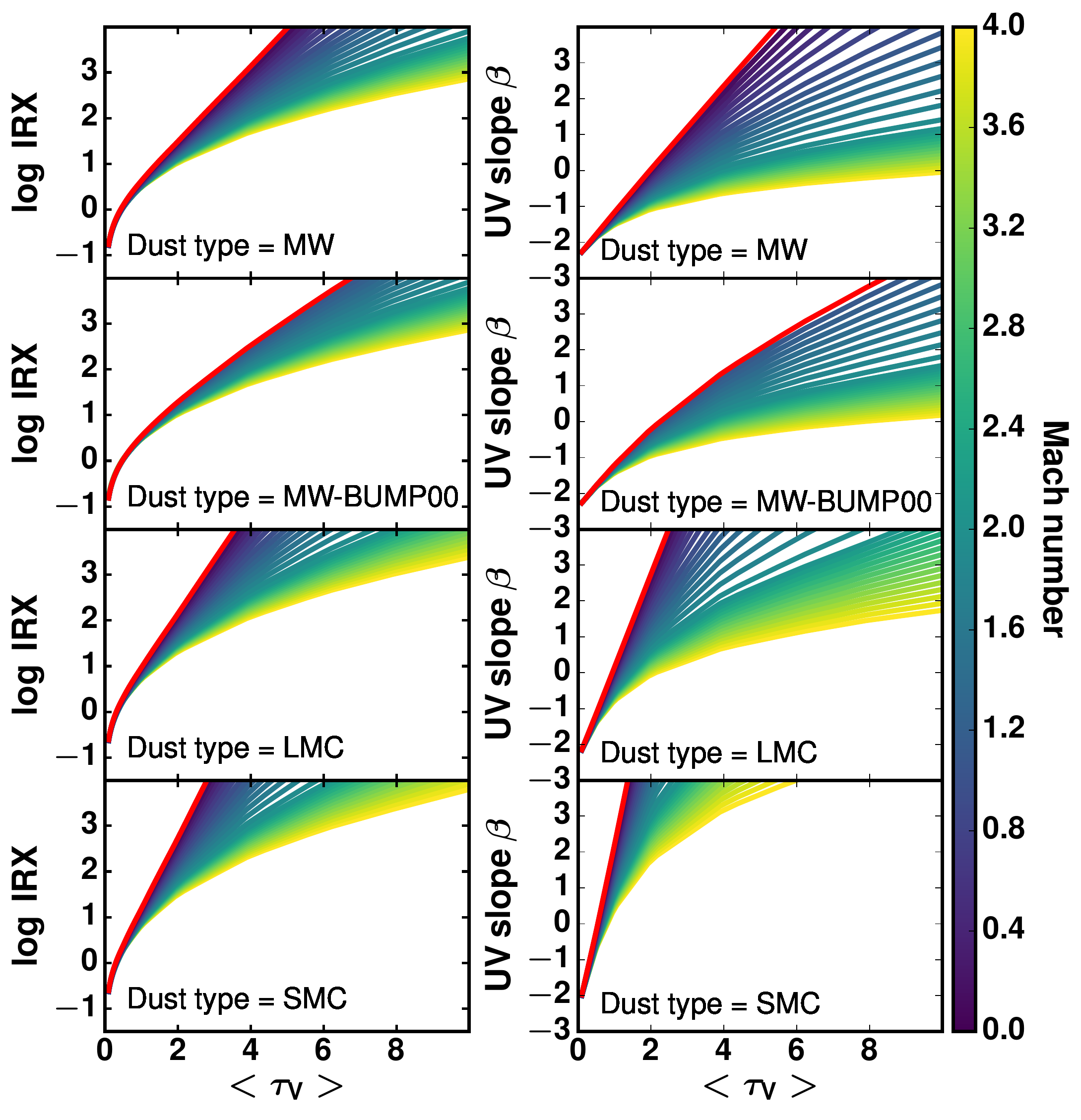}
\caption{The UV-slope $\beta$ and infrared excess (IRX) as a
  function of mean homogeneous optical depth in the V-band and Mach
  number for different types of dust attenuation curves. The red lines
  show $\beta$ and IRX for a dust screen without turbulence (i.e.,
  $\mathcal{M}=0$). As the level of turbulence increases, both $\beta$ and IRX decrease.
\label{fig:break_down_Mach}}
\end{figure}
\begin{figure*}
\includegraphics[width = 0.65\hsize]{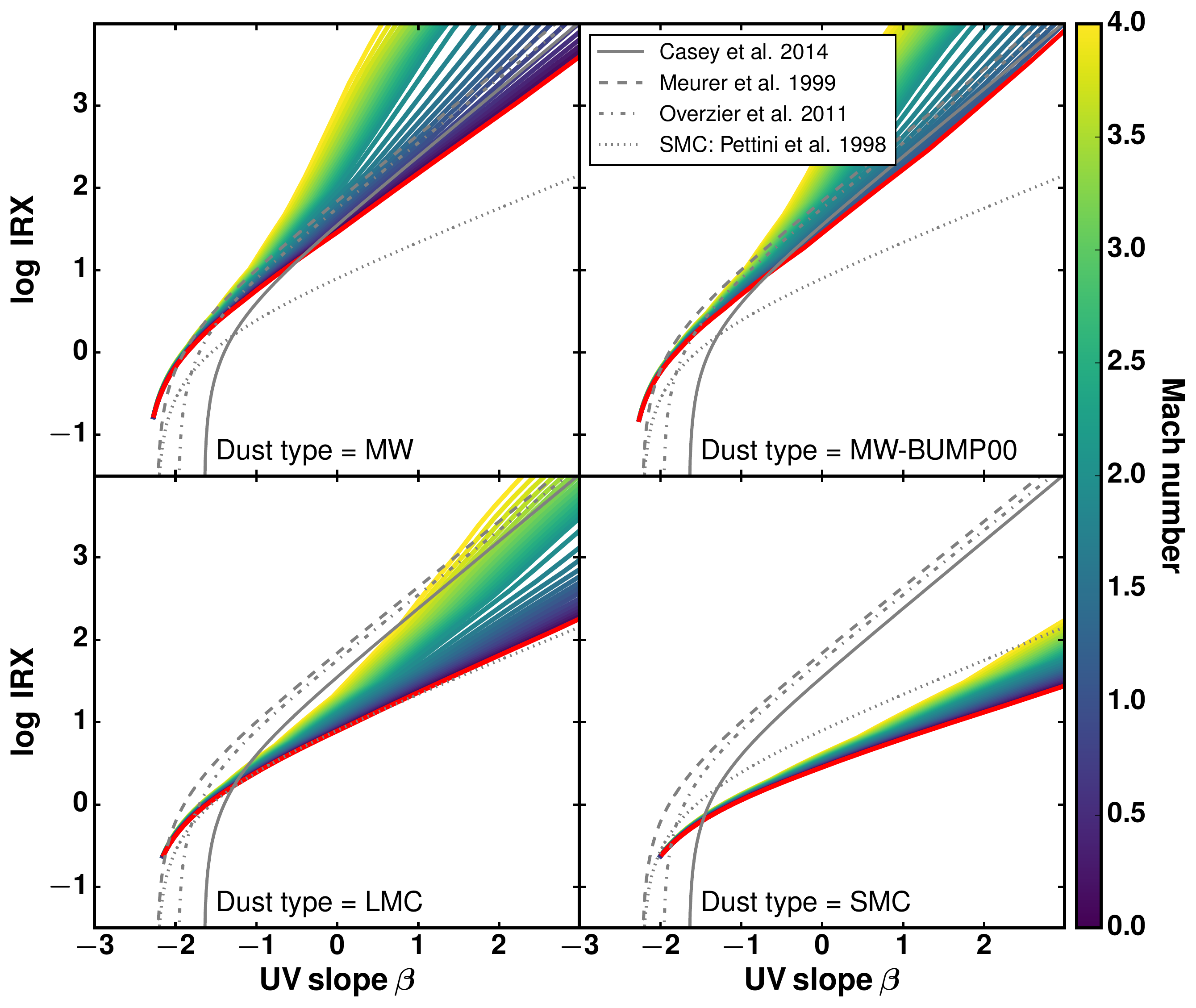}
\caption{The \IRXB relation for varying levels of turbulence within
  the screen of dust. The relations are plotted for different dust
  types. The red lines
  show the \IRXB relation for a dust screen without turbulence (i.e.,
  $\mathcal{M}=0$). Turbulence within the
  dust screen can significantly alter the location of a galaxy within
  the \IRXB plane.
\label{fig:Mach_beta_IRX}}
\end{figure*}

\subsection{Calculating $\beta$ \& IRX}
The UV-slope $\beta$ is computed assuming that the (attenuated)
stellar emission can be
described by  a power-law between wavelength and flux,  with $F_\lambda \propto
\lambda^{\beta}$. The UV-slope $\beta$ is approximated by a log-linear fit in the wavelength range 1230
-- 3200 \Angstrom. This wavelength range corresponds to UV radiation from O and B
stars and ensures a broad wavelength coverage around the 2175 \Angstrom
feature, if present.

The infrared excess (IRX) is computed as
\begin{equation}
\text{IRX} = \frac{L_{\text IR}}{L_{1600\,\Angstrom}},
\end{equation}
where $L_{1600\,\Angstrom}$ is the UV luminosity at 1600 \Angstrom and
$L_{\text IR}$ is the infrared luminosity. We assume
that all absorbed stellar emission $L_{\text abs}$ is re-emitted in the
infrared, which gives an infrared luminosity of $L_{\text IR} =
L_{\text abs}$.

\section{Results}
\label{sec:analysis}
In this Section we vary the properties of the stellar population and
the screen of dust to explore how such variations change the \IRXB
relation of dust attenuation. We will vary the stellar age and
dust attenuation curve (Sec. \ref{sec:age_curve}), introduce turbulence in
the dust screen (Sec. \ref{sec:Mach}), vary the covering fraction of
the dust screen (Sec. \ref{sec:cov_frac}), mix the stars in between
the dust (Sec. \ref{sec:dust_distribution1}), introduce a
two-component dust model \citep[Sec. \ref{sec:two_comp}]{Charlot2000},
and explore how observational effects change the measured \IRXB relation
(\ref{sec:observational_effects}). Throughout this section we compare our
results to four \IRXB relations from the literature: The seminal
\citet{Meurer1999} relation for local starburst;\footnote{The
  \citet{Meurer1999} \IRXB relation is based on a FIR luminosity
  covering the wavelength range 
  40--120 $\mu$m, rather than the entire IR wavelength regime. \citet{Meurer1999} find that these are typically
  different by a factor of $\sim$1.4.} a refit to the
galaxies in \citep{Meurer1999} by \citet{Overzier2011} based on
\emph{Galaxy Evolution Explorer} (\Galex) data; a fit by
\citet{Casey2014} to local galaxies covering a large range in
SFRs; and the \IRXB relation for SMC dust based on \citet{Pettini1998}. 

\subsection{Varying dust type and stellar population age}
\label{sec:age_curve}
We show the \IRXB relation for the four different types of dust
attenuation curves explored in this work (MW, MW-BUMP00, LMC, and SMC)
for stellar populations with ages running from 1 to 100 Myr in Figure \ref{fig:fiducial_beta_IRX}. The SMC relation is flatter than the MW
relation, with the LMC relation lying in between. Due to the elevated level of attenuation
in the UV regime for the SMC and LMC attenuation curves (see Figure
\ref{fig:attenuation_law}), the slope $\beta$ will be steeper compared
to MW dust types. As the optical depth of the dust slab
increases, the FUV emission is relatively more absorbed than the NUV. This moves the \IRXB relation to the right compared to the
MW type attenuation curves and causes the flattening in the \IRXB relation from MW to SMC
dust types. These results immediately
demonstrate that simply by changing the dust-type the relation between
UV-slope and infrared excess can change significantly. There is hardly
any difference between the MW and MW-BUMP00 \IRXB relation, which
suggests that the effect of
the 2175 \Angstrom on the location of galaxies in the \IRXB plane is
minimal.\footnote{We will show in Section \ref{sec:observational_effects} that
  the 2175 \Angstrom bump can contaminate a poor photometric sampling
  of the UV spectra of galaxies and affect the reconstructed \IRXB relation.}

There is a clear trend in the \IRXB relation as a function
of the age of the stellar populations. With age, the UV part of a
stellar spectrum becomes redder, naturally shifting the \IRXB
relation to higher values of the UV-slope $\beta$. 

The simple model presented in this work reproduces the
observed relations very well. The shapes of the theoretical \IRXB relation are similar to
the observed relations. The MW models for a stellar population
with an age of 10 Myr are very close to the \citet{Casey2014} fit to
local galaxies. The \citet{Meurer1999} and \citet{Overzier2011} fit to
local starbursts it typically located above our model results (except
for populations with ages of $\sim 1$ Myr). The LMC and
SMC curves for a population of 10 Myr are both close to the \IRXB
relation for SMC dust in the literature \citep{Pettini1998}. The
match between our predicted and the observed \IRXB relations suggests
that the observed relations are
consistent with a simple screens of dust
in front of a relatively young population of stars. 
\begin{figure}
\centering
\includegraphics[width = 0.9\hsize]{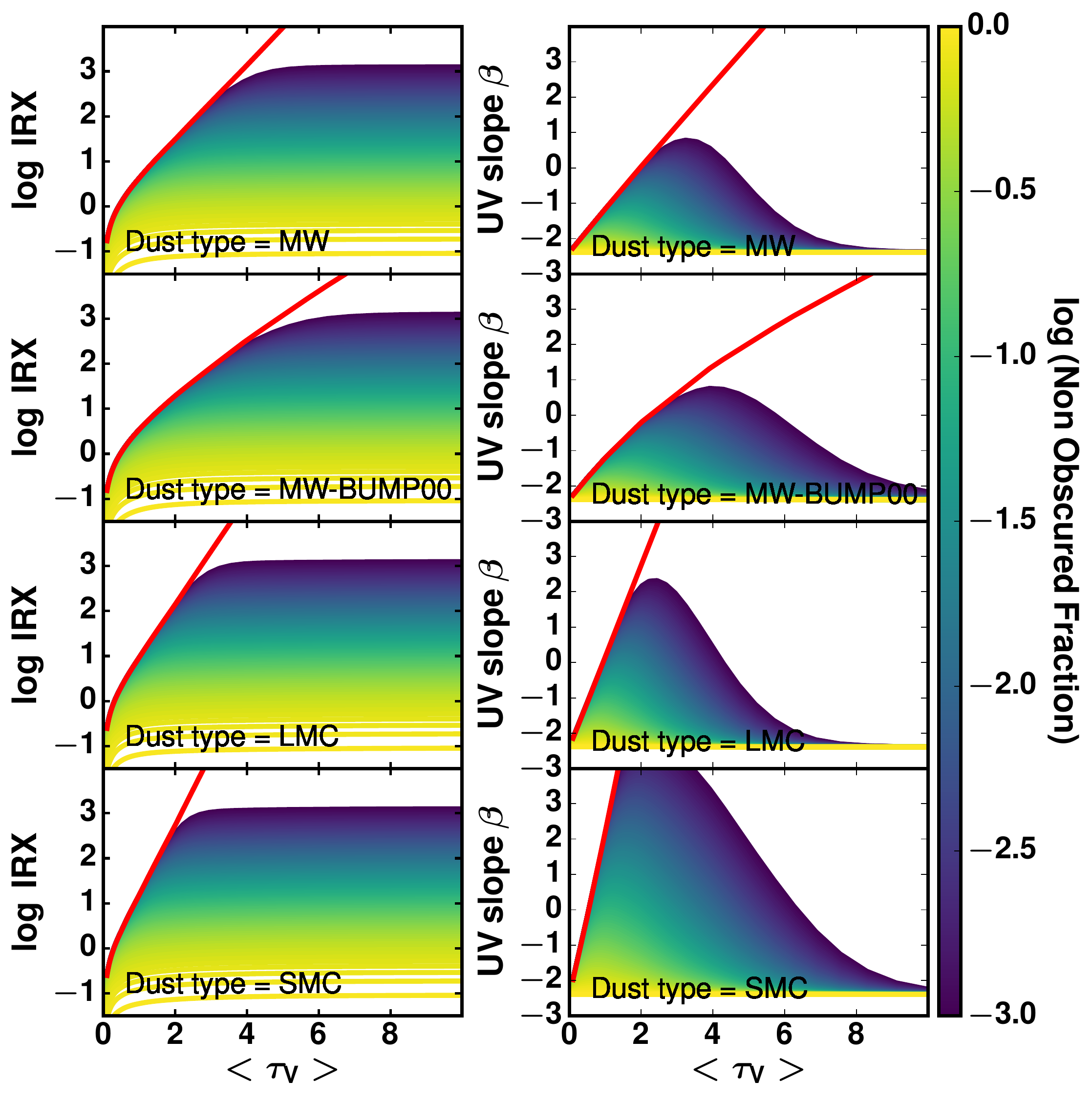}
\caption{The UV-slope $\beta$ and infrared excess (IRX) as a
  function of mean homogeneous optical depth in the V-band and the
  obscuration fraction of the stellar emission for different dust attenuation curves. The red lines
  show $\beta$ and IRX for a dust screen with a non-obscured fraction
  of zero (i.e., all the emission passes through a dust screen). As the
  non-obscured fraction increases, both $\beta$ and IRX decrease.  Most
  intriguingly, a small non-obscured fraction can already
  lead to blue UV-slopes for high-values of $<\tau_{\text
    V}>$. \label{fig:break_down_comp}}
\end{figure}
\begin{figure*}
\includegraphics[width = 0.65\hsize]{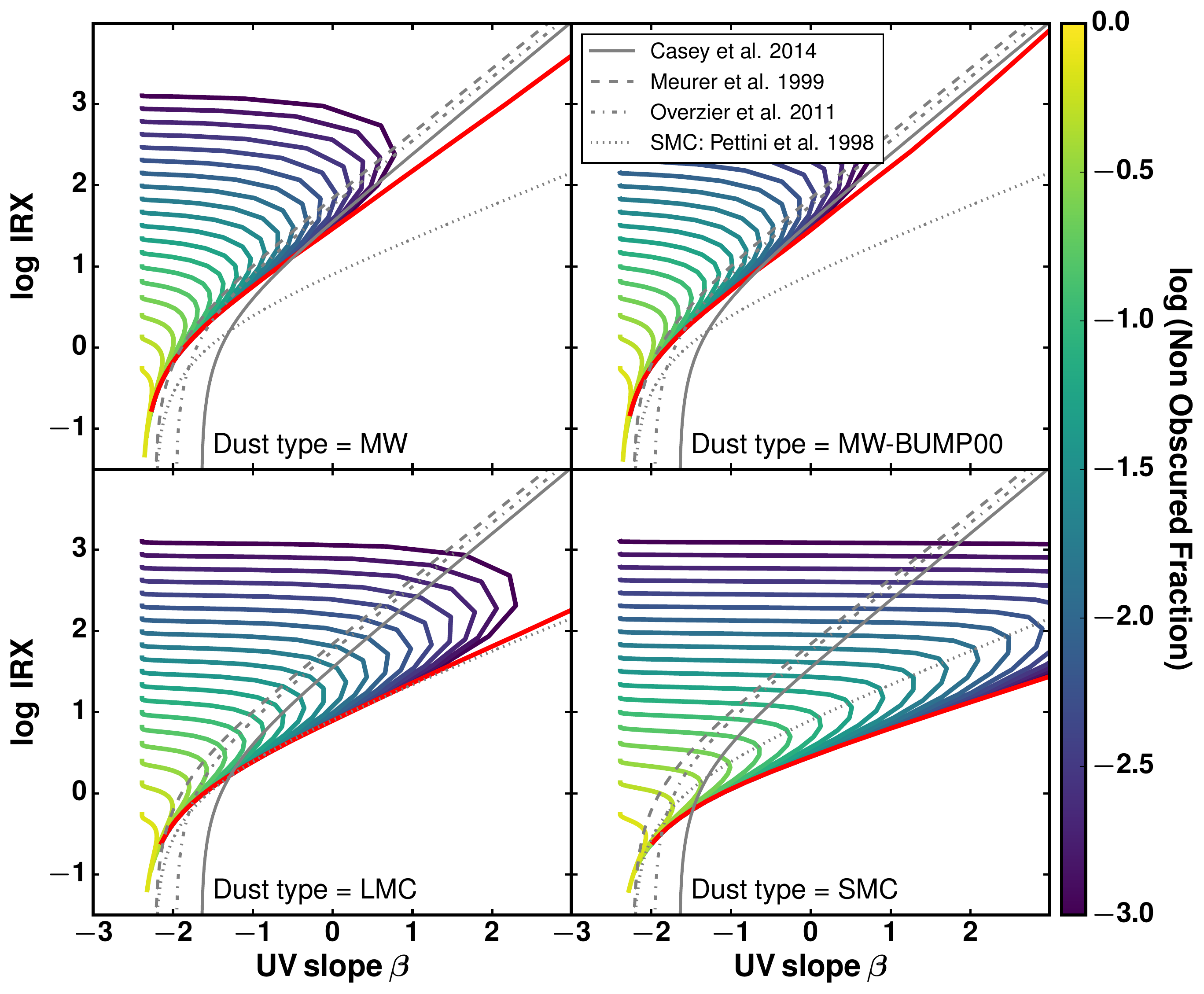}
\caption{The \IRXB relation when a fraction of the stellar light
  is not obscured. The relations are plotted for different dust
  types. The red lines
  show the \IRXB relation for a dust screen with an non-obscured
  fraction of zero (i.e., all the emission passes through a dust screen). A small fraction of
  non-obscured emission can completely alter the location of a galaxy
  within the \IRXB plane, with high IRX and blue UV-slope $\beta$.
\label{fig:2comp_beta_IRX}}
\end{figure*}

\subsection{A turbulent sheet of dust}
\label{sec:Mach}
A small level of turbulence within a dust screen can locally increase
or decrease the column of dust the stellar emission has to travel
through. This turbulence can for instance be caused by
  shocks or shear motions within spiral arms. Although the volume weighted mean optical depth due to absorption remains
the same, locally the optical depth varies. Because the absorption
of light scales exponentially with optical depth, the net amount of
light passing through the screen of dust will vary. This was
initially explored by \citet{Fischera2003}, who showed that
with an increasing level of turbulence more light passes through a
screen of dust. The authors found that turbulence drives dense clumps
of dust with high optical depth where no light passes through. This is
balanced by large areas with low optical depth where
almost all the stellar emission passes through. This was explored
further by \citet{Seon2016} who reached similar
conclusions and demonstrated that the shape of the net attenuation
curve (by averaging over all the sight lines) changes with increasing level of turbulence. A change in shape
of the attenuation curve will naturally affect the UV-slope
$\beta$ and IRX.

We show how $\beta$ and IRX change as a function of $<\tau_{\text V}>$ and Mach number in Figure \ref{fig:break_down_Mach}. Both $\beta$
and IRX decrease when the level of turbulence increases. IRX
decreases due to the overall level of absorption decreasing. The UV
emission escaping the dust screen is dominated by emission from regions with low
optical depth. In these regions the attenuated stellar emission is closer to the
original unattenuated spectrum than for a homogeneous screen of
dust. This naturally decreases the UV-slope $\beta$. The same trend
occurs for the four different dust types.

We show the \IRXB relation for varying levels of turbulence in Figure
\ref{fig:Mach_beta_IRX}. The \IRXB relation shifts above the
relation for a homogeneous dust screen with increasing level of
turbulence. This effect is stronger for MW dust types than for LMC and
SMC types of dust. In other words, this effect becomes less important
for attenuation curves with a strong FUV component. The scatter in IRX
increases as a function of UV slope $\beta$.

\begin{figure*}
\includegraphics[width = 0.65\hsize]{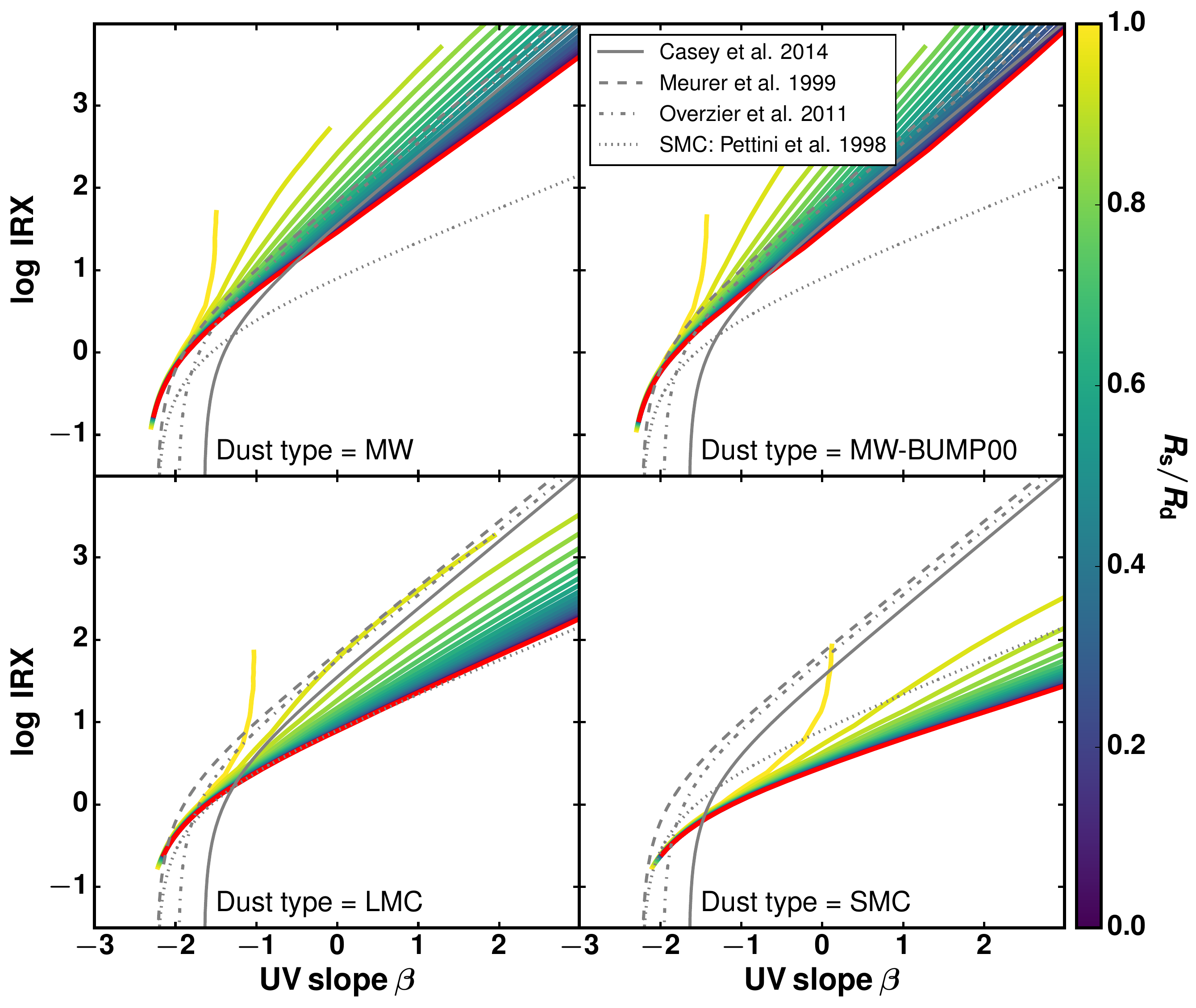}
\caption{The \IRXB relation when the young stars are distributed
  within the dust screen (rather than placed before the dust screen). The relations are plotted for different dust
  types. The red lines
  show the \IRXB relation for the scenario where $R_{\rm s}/R_{\rm d}
  =0$ (i.e., the dust screen is located between the stars and the observer). Mixing the stars between the dust lowers IRX
  values, makes the attenuated spectra bluer, and causes offsets above
  the $R_{\rm s}/R_{\rm d} = 0$ \IRXB relation.
\label{fig:dustDist_beta_IRX}}
\end{figure*}

\begin{figure*}
\includegraphics[width = 0.65\hsize]{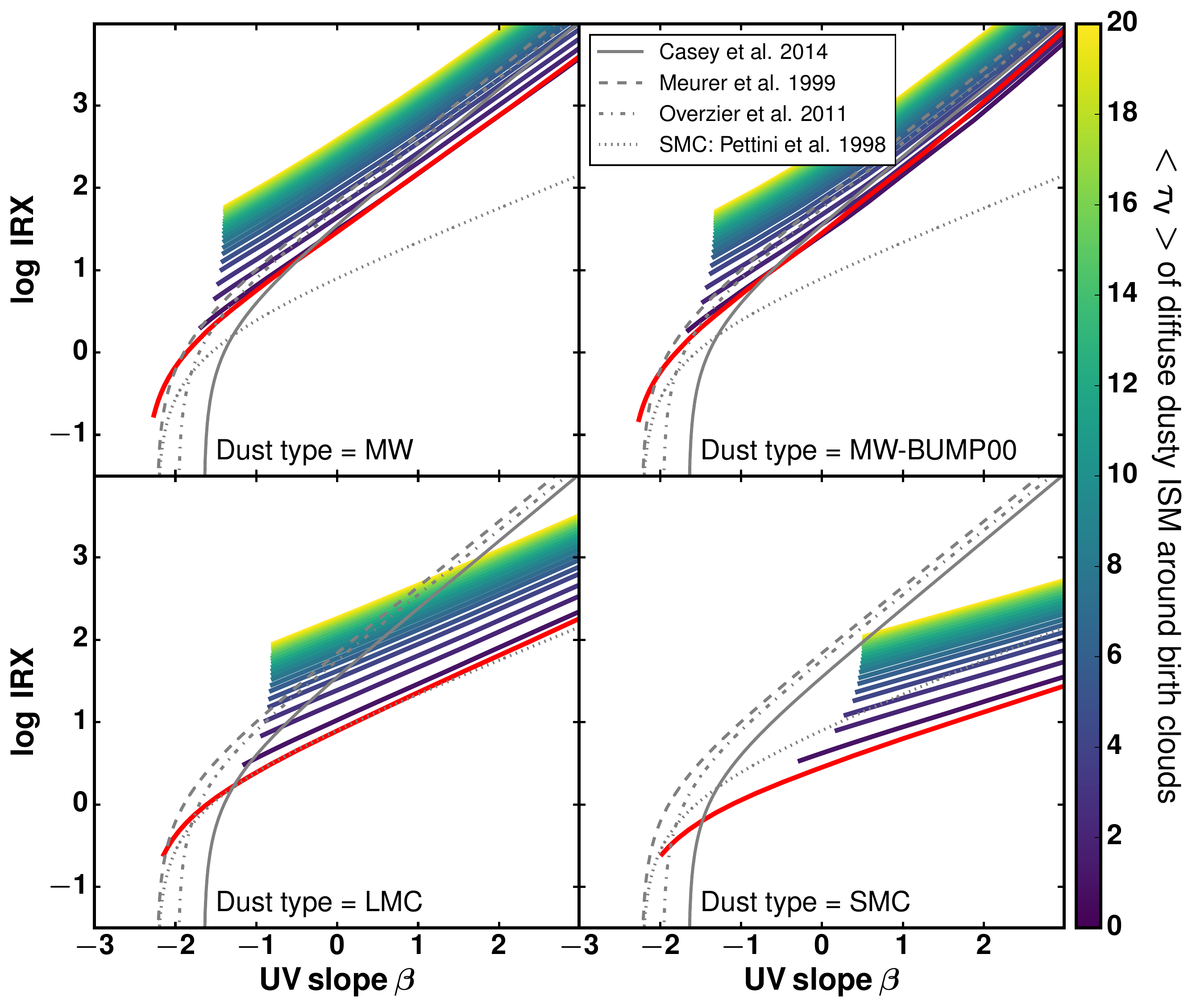}
\caption{The \IRXB relation when we assume that birth clouds are
  homogeneously distributed within a diffuse dusty ISM
  \citep[two-component dust model;][]{Charlot2000}. The stellar emission passes
  through two screens. First a screen with $R_{\rm s}/R_{\rm d}=0$,
  representing the birth clouds of stars. The second screen
  representing the diffuse dusty ISM has $R_{\rm s}/R_{\rm d} =1$,
  a homogeneous mixing of photon sources (the
  emission escaping from birth
  clouds) and dust. In this Figure the optical depth $<\tau_{\rm V}>$ of both screens
  are varied. The colour coding corresponds to fixed mean homogeneous optical depths of the screen of
  dust representing the diffuse dusty ISM. An increase in IRX along a
  line with fixed colour corresponds to an increase in the homogeneous
  optical depth of the birth cloud. The red lines
  show the \IRXB relation for the scenario where the second screen
  representing the diffuse dust ISM has $<\tau_{\rm V}>=0$, i.e.,
  stellar emission only passes through the birth cloud.  The relations are plotted for different dust
  types. Homogeneously
  distributing the birth clouds of stars within a dusty ISM causes
  variations in the \IRXB relation towards bluer values of $\beta$ for
  a fixed IRX.
\label{fig:IRX_beta_two_comp}}
\end{figure*}

\begin{figure*}
\includegraphics[width = 1.0\hsize]{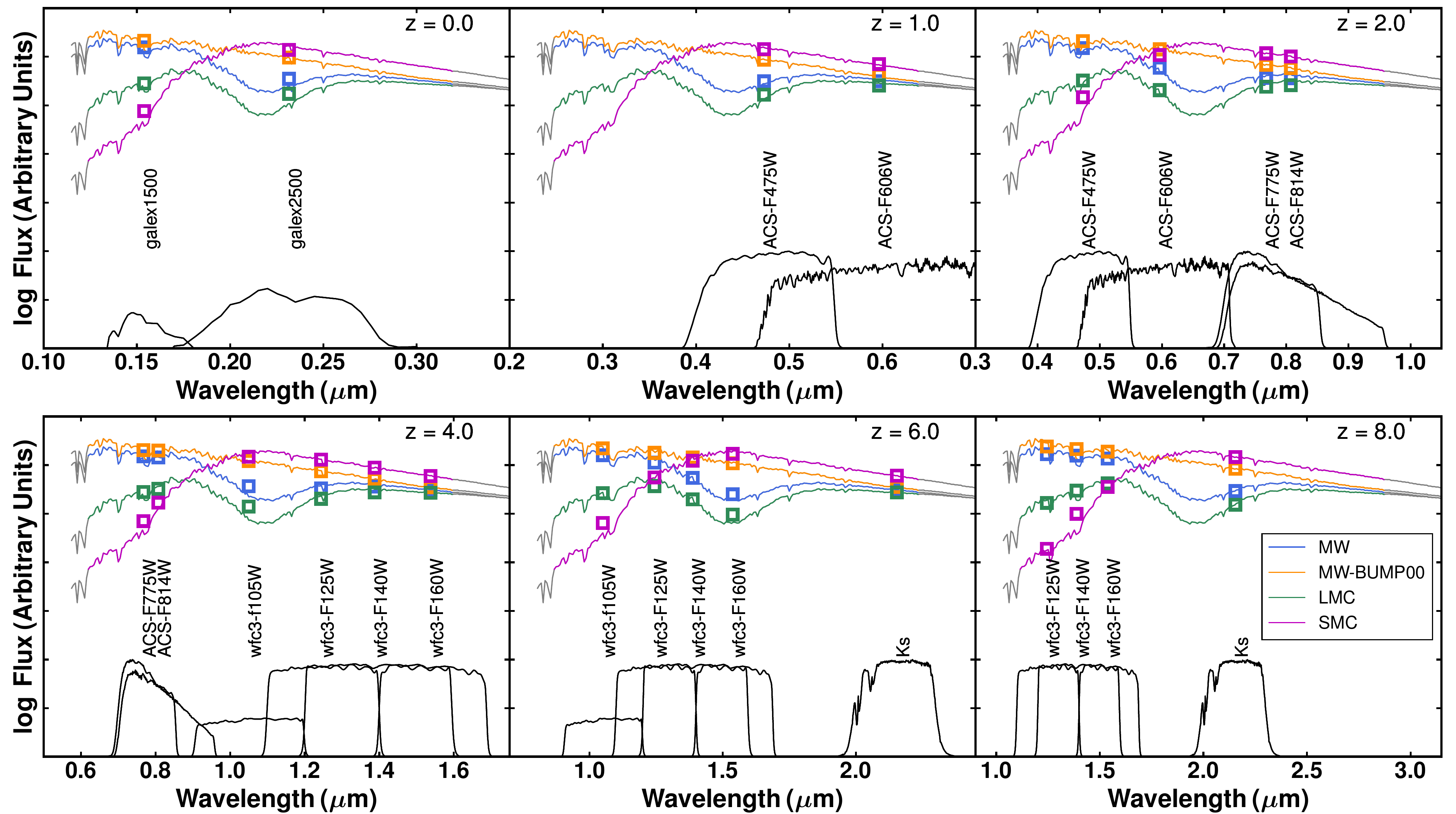}
\caption{The coverage of the UV rest-frame wavelength regime by
  different filters for redshifts running from $z=0$ to $z=8$, when
  adopting the four different dust-types in this study. The
  photometric sampling 
  includes the HST filters that are part of the 3D-HST survey
  \citep{Skelton2014}, on top of \Galex, and a
  Ks-band filter. The coloured
  lines mark the attenuated spectra (from rest-frame 1230 to 3200 \Angstrom)
  assuming a V-band optical depth of
  $<\tau_{\text V}>=1$. The relevant filters at each redshift and their
  transmission curves are plotted below the stellar spectra. The
  photometric datapoints are marked as coloured squares, where the
  colours correspond to the attenuated spectra for different dust
  attenuation curves. The stellar
  spectra come from a solar-metallicity single-burst population with
  an age of 10 Myr. The  sampling of the filters does not
  always give a good representation of the original spectrum (e.g., the poor sampling of the absorbed FUV regime at $z=1$ for the SMC
  dust type).
\label{fig:observed_spectra_3DHST}}
\end{figure*}

\subsection{The covering fraction of the dust}
\label{sec:cov_frac}
In the previous Section we explored the effects of turbulence within
the dust screen on the \IRXB relation, where turbulence led to a locally
elevated or decreased optical depth. A more extreme version of this is
a patchy screen, i.e. a screen with holes in it. If the screen is
patchy and thus has a dust covering fraction less than one (the screen has holes), some fraction of the stellar emission can escape the
screen without ever being absorbed or scattered. 

In Figure \ref{fig:break_down_comp} we show how UV-slope $\beta$ and
IRX evolve as a function of mean homogeneous optical depth in the
V-band of the dust screen and the fraction of non-obscured emission
(i.e. the fraction of emission not passing through a dust
screen). A non-obscured fraction
of zero means that all of the stellar emission passes through a dust
screen, whereas a non-obscured fraction of one means that none of the
emission passes through a dust screen (in other words, there is no
screen of dust between the photon source and the observer). We find that both the UV-slope $\beta$ and the infrared excess
IRX decrease with an increased fraction of non-obscured
emission. When less emission is obscured, the fraction of absorbed
emission automatically decreases, which yields lower values of
IRX. Similarly, the UV emission from the non-obscured regime will
dominate over the UV emission from the obscured regime, resulting into
a lower value for $\beta$. Interestingly, when a non-obscured
component is present the UV-slope $\beta$ increases and then decreases
again as a function of mean homogeneous optical depth. The turning
point marks the mean homogeneous optical depth where UV emission from the non-obscured regions
starts to dominate over the UV emission from the obscured regions. IRX
remains almost constant once the optical depth is larger than the optical depth
corresponding to the turning point in $\beta$. At these optical depths the slab
of dust is so optically thick that effectively all the stellar
emission is absorbed and re-emitted in the IR. From this point on, the IR luminosity 
thus remains almost constant as the optical depth increases. The UV emission is
completely dominated by the UV emission from the non-obscured region,
and also remains constant. This results in a constant value for IRX. It is
worth noting that if the non-obscured fraction of the stellar emission
is very small, the UV emission escaping the system
may be too faint to observe.

We show the \IRXB relation as a function of non-obscured fraction for
four different dust types in Figure \ref{fig:2comp_beta_IRX}. When a
non-obscured component is present, the \IRXB relation first follows
the relation for a dust screen completely obscuring the stellar
emission. However, the relation then turns backwards towards lower values
of $\beta$ while still increasing IRX. As before, the turning points marks
the mean homogeneous optical depth where UV emission from non-obscured
regions starts to dominate over the UV emission from obscured regions (Figure
\ref{fig:break_down_comp}). A very low non-obscured
fraction (less than 1\%) in combination with a large optical depth
result in a scatter in IRX up to 2--3 dex for a given UV slope
$\beta$.

\subsection{Mixing of stars and dust}
\label{sec:dust_distribution}
\subsubsection{Mixing of stars and dust within a birth-cloud}
\label{sec:dust_distribution1}
 We have so far assumed that the screen of dust is
located between the stars and the observer (i.e., $R_{\rm
  d}/R_{\rm s} = 0$). We will now explore the effects of mixing stars
with the dust (varying $R_{\rm
  d}/R_{\rm s} $) on the \IRXB relation in Figure
\ref{fig:dustDist_beta_IRX}. The \IRXB relation shifts towards higher
values of IRX and lower values for UV-slope $\beta$ with increasing $R_{\rm
  d}/R_{\rm s} $. As $R_{\rm
  d}/R_{\rm s} $ increases, some of the stars will see a smaller
column of dust (since they are located towards the edge of the screen
of dust facing the observer). As a result, the total absorption of starlight and
consequently IRX both decrease. Because some of the stars are less
attenuated the spectrum will also remain bluer, decreasing
the UV-slope $\beta$. This effect is strongest for a homogeneous
mixing of dust and stars ($R_{\rm
  d}/R_{\rm s} =1 $).  For this scenario there will always be stars
whose emission is not attenuated at all, which rapidly results in 
lower values for IRX and $\beta$.

\subsubsection{Dusty birth clouds within a diffuse dusty ISM}
\label{sec:two_comp}
A more realistic representation of the mixing of stars and dust is the
two-component model put forward by \citet{Charlot2000}. In this model the
birth clouds of stars are homogeneously distributed within a diffuse
dusty ISM. To represent this we assume that the emission from stars is exposed to two
screens of dust. First, a screen of dust with $R_{\rm
  d}/R_{\rm s} =0$, representing the birth cloud surrounding the young
stars. The attenuated stellar emission that escapes the first screen passes trough a second dust screen with $R_{\rm
  d}/R_{\rm s} =1$. This means that the birth clouds are homogeneously
distributed within a second screen of dust representing the diffuse
dusty ISM. The resulting \IRXB relations
are presented in Figure \ref{fig:IRX_beta_two_comp}. The colour coding
in this Figure corresponds to  fixed optical depths for the screen of dust
representing the diffuse dusty ISM. An increase in IRX along a line
with fixed colour corresponds to an increase in the optical depth
of the screen representing the birth clouds.

The \IRXB relation moves towards bluer UV colours $\beta$ at fixed IRX
as the optical depth in the screen of dust representing the
diffuse dusty ISM increases. This is driven by variations
in optical depth seen by the stellar emission that escapes the birth
clouds. For some of the birth
clouds the emission that reaches the observer had to travel through the entire slab representing the
diffuse dusty ISM. The stellar emission is
further absorbed and contributes to the infrared emission (and thus
drives IRX to higher values). Other birth clouds will be located at
the observer side of the slab of dust representing the diffuse dusty
ISM. For these clouds the escaped stellar emission (almost) directly travels to the
observer without additional absorption. This emission is the
dominant source of the UV emission that reaches the observer and will
control the UV slope $\beta$. While IRX is weighted towards stellar emission
that is further absorbed, $\beta$ is weighted towards stellar emission that is not
further absorbed. As a result, $\beta$ is relatively blue
for a given IRX.

As becomes apparent from Figure \ref{fig:IRX_beta_two_comp}, an
increase in $<\tau_{\rm V}>$ for the screen of dust representing the diffuse
dusty ISM results in a minimum value for the UV slope $\beta$. This
minimum represents the scenario where the optical depth of the birth
cloud equals zero (the stars are not located within birth clouds
anymore). For this scenario the two-component model is identical to the
scenario explored in Section  \ref{sec:dust_distribution1} with
$R_{\rm s}/R_{\rm d} = 1$.

Changes in the \IRXB relation due to the two-component model become
significant when the diffuse dusty ISM has a homogeneous mean optical
depth of $<\tau_{\rm V}> \sim  4$, or higher. These environments
should only be relevant for relatively dust-rich systems such as DSFGs.

\begin{figure*}
\includegraphics[width = 0.65\hsize]{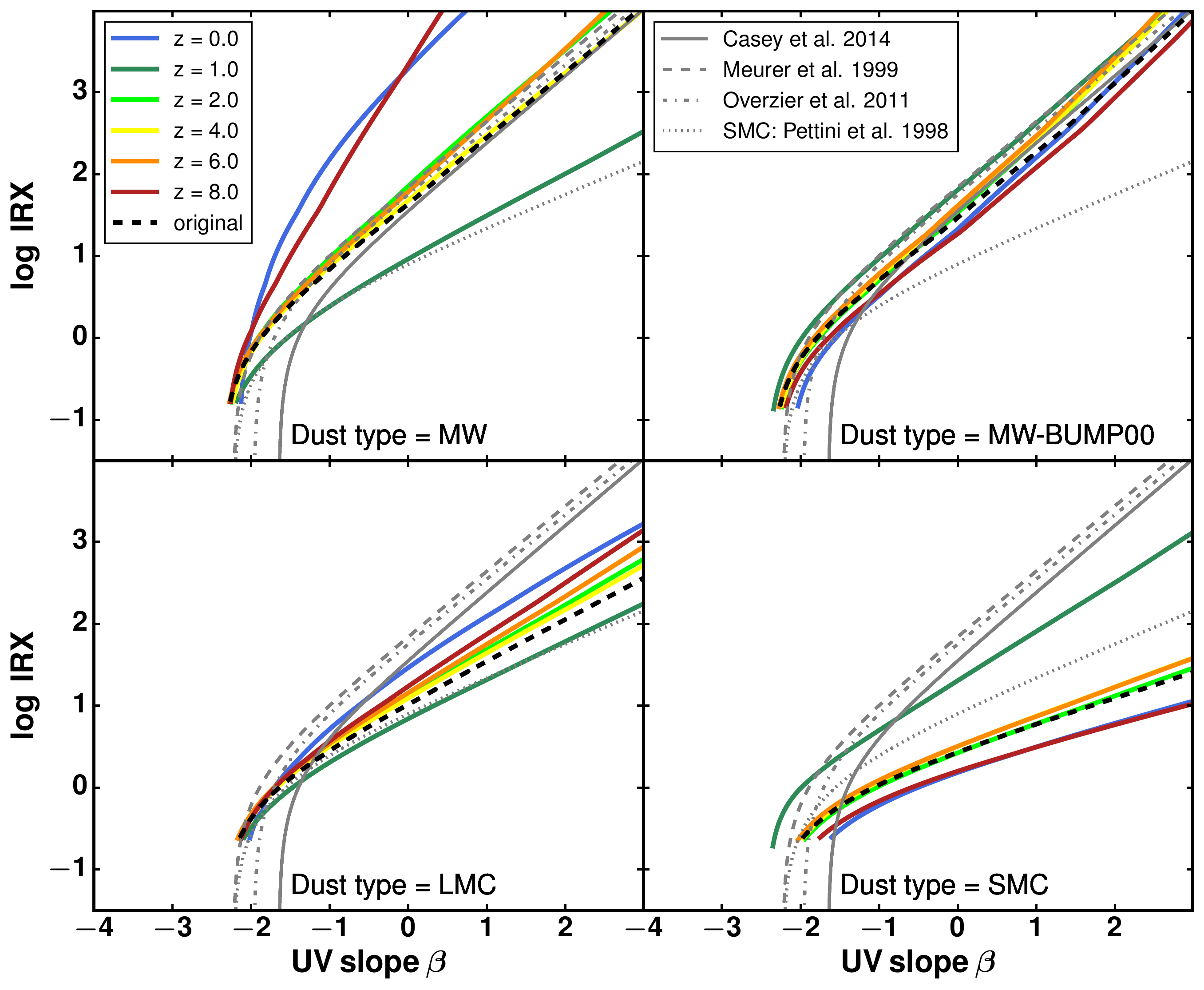}
\caption{The \IRXB relation based on a fit to photometric
  datapoints for different dust types. The
  photometric sampling   includes the HST filters that are part of the 3D-HST survey
  \citep{Skelton2014}, on top of \Galex filters and a
  Ks-band filter. Photometric datapoints were obtained by convolving the
  redshifted stellar spectrum with telescope filters corresponding to
  the rest-frame wavelength range $1230<\lambda<3200$ \Angstrom.  The black dashed lines correspond to the original
  \IRXB relation derived from a direct fit to the attenuated
  spectrum. The 
  sampling of telescope filters of the UV wavelength range 
  alters the location of galaxies in the \IRXB plane for
  dust types with a steep UV-slope (SMC dust) or UV absorption
  features (2175 \Angstrom bump; MW and LMC dust). 
\label{fig:observed_beta_IRX}}
\end{figure*}

\subsection{Deviations from  \IRXB relation due to observational effects}
\label{sec:observational_effects}
Up to this point the UV-slope $\beta$ was calculated by a fit to the
attenuated stellar spectrum. Unfortunately, observations are often
limited to photometric datapoints (especially at high redshifts) and $\beta$ is then computed based
on a fit to  photometric sampling of the UV emission
\citep[e.g.,][]{Casey2014,Capak2015,Puglisi2016,Smit2016, Reddy2017}. The 1600 \Angstrom UV luminosity that
enters the IRX calculations is estimated by an interpolation
between the photometric datapoints. To mimic this, in this Section we calculate the
UV-slope $\beta$ and 1600 \Angstrom UV luminosity based on a fit to
photometric datapoints of the attenuated stellar emission from $z=0$ to $z=8$. Photometric datapoints were obtained by convolving the
redshifted attenuated stellar light with telescope filters probing the
rest-frame UV wavelength range ($1230 < \lambda
< 3200 \Angstrom$). Depending on the redshift of interest and
combination of filters, the UV-slope
$\beta$ is calculated based on 2 up to 6 (or even more) photometric datapoints. 

Figure
\ref{fig:observed_spectra_3DHST} gives an example of the coverage of the UV rest-frame
wavelength regime by different filters at different
redshifts. We use the same set of filters that were used for the 3D-HST
survey \citep{Skelton2014}, combined with \Galex, and a Ks-band filter. This
Figure immediately demonstrates that in some cases the power-law fit
to the photometric datapoints may suffer from a poor sampling of
strong FUV absorption (for example at $z=1$ for an
SMC dust attenuation curve) or
a contamination of the fit by the 2175 \Angstrom bump (for example at
$z=0$, $z=1$, and $z=8$).\footnote{Observational constraints on the
  2175 \Angstrom bump indicate that
  this bump is typically not as strong in extragalactic sources as in
  our own Galaxy \citep[e.g.,][]{Kriek2013}} A poor sampling or contamination can result in a
poor estimation of the UV-slope $\beta$ and the UV luminosity,
introducing a systematic error in the \IRXB relation \citep{Kriek2013}.\footnote{Besides a poor fit to
the rest-frame UV wavelength regime, additional uncertainty may be
introduced when estimating the total IR flux from galaxies based on a
limited sampling of the FIR and sub-mm SED or incorrect assumptions on
the temperature of the dust \citep{Bouwens2016,Narayanan2017}. We ignore these
effects in this work, but caution the reader to be aware of them.}

We show the reconstructed \IRXB relation based on fits to photometric
datapoints for different redshifts in Figure
\ref{fig:observed_beta_IRX}. The original \IRXB relation based on a
direct fit to the attenuated spectra is shown as a black dashed line. The \IRXB
relation based on a fit to photometric datapoints can significantly
differ from the original \IRXB relation. This is especially evident
for MW dust at $z=0$, $z=1$, and $z=8$, driven by contamination of the
fit by the 2175 \Angstrom bump. At $z=0$ and $z=8$ the poor sampling
suggests that the attenuated stellar emission is very blue for a given
IRX. At $z=1$ the resulting \IRXB relation is similar to the SMC \IRXB
relation \citep{Pettini1998}, even though a MW dust attenuation curve
was employed. The same effects are seen for the LMC dust
attenuation curve, though less prominently. There is a clear offset
between the \IRXB relation based on a fit to photometric datapoints and a
fit to the attenuated stellar spectrum  for the SMC dust attenuation
curve at $z=1$. This is driven by a poor sampling of the FUV part
of the rest-frame spectrum. There is hardly any difference between the
'observed' and original \IRXB relation for the MW-BUMP00 dust
attenuation curve. The MW-BUMP00 attenuation curve doesn't have any
strong features and the attenuated stellar spectrum can
easily be fit even with only two photometric datapoints. 

Additional coverage of the rest-frame NUV wavelength range at $z=0$
and $z=8$, as well as additional coverage of the rest-frame FUV
wavelength regime at $z=1$ would improve the reconstruction of the
\IRXB relation. We explore other filter combinations in Appendix
\ref{sec:appendix} and in some cases indeed find a better reconstruction
of the \IRXB relation. We list suggested filter combinations as a
function of redshift that reliably reconstruct the UV slope
$\beta$ of the \IRXB relation in Table \ref{tab:filter_combs}.

\section{Discussion}
\label{sec:discussion}
\begin{figure}
\includegraphics[width = \hsize]{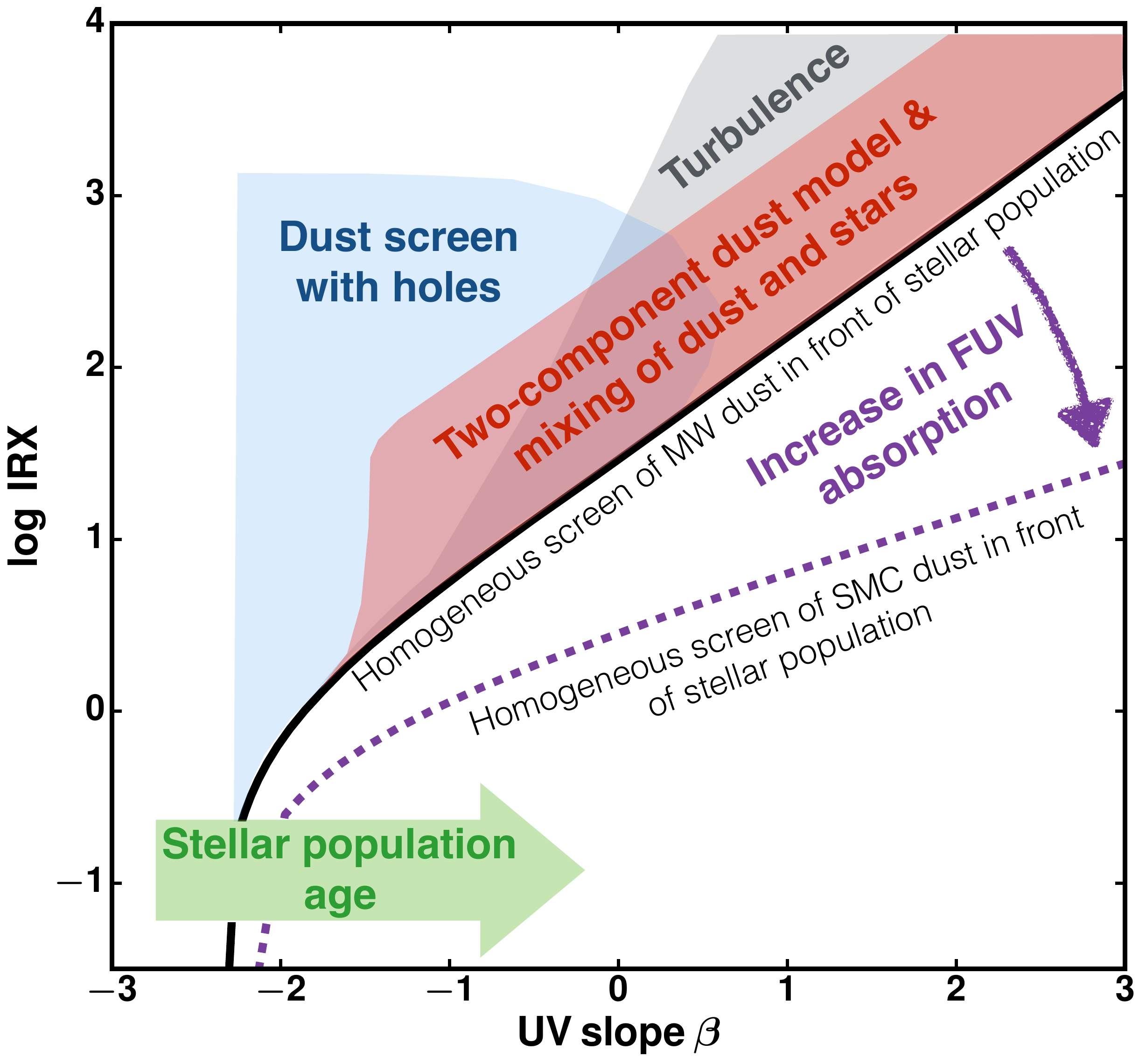}
\caption{A schematic figure summarising how different physical processes
  affect the \IRXB relation. A patchy screen, turbulence, the mixing of dust and
  stars, and a two-component dust model yield deviations above the MW \IRXB relation, whereas age and
  an increased FUV absorption component yield deviations below the
  \IRXB relation.
\label{fig:schematic}}
\end{figure}

In this work we have presented a simple model that puts a screen of
dust in front of a young stellar population to study how varying
properties of the dust screen and the stellar population cause
variations in the \IRXB dust attenuation relation of
galaxies. It is worthwhile to separate deviations in two
different classes, above and below the original \IRXB relation, or in
other words, deviations towards bluer and redder UV colours at fixed IRX. Older stellar populations move the \IRXB
relation to redder UV colours. An attenuation curve with a strong FUV
component flattens the relation, which results in deviations towards
redder UV colours compared to the MW dust attenuation
curve. An increased level of turbulence, on the other hand, results in bluer UV
slopes at fixed IRX.  A small region of unobscured stellar emission
(up to a few per cent), the mixing of dust and stars, and a
two-component dust model (where dusty birth clouds are located within
a diffuse dusty ISM) also
results in deviations towards blue UV colours for a given IRX. We schematically depict these effects in
Figure \ref{fig:schematic}. In
the remaining of this Section we will first focus more on the physics
behind the \IRXB relation. We then discuss the origin of variations in the \IRXB
relation for different observed classes of galaxies. We finish by
discussing the prospects of approximate analytical equations for the
\IRXB relation and what future observations can better constrain the
\IRXB relation as a function of galaxy type and cosmic epoch.

\subsection{The physics behind the \IRXB relation}
The \IRXB
relation for a uniform screen of dust has a very distinct shape, best described by an asymptotic
power law. The individual components (the power law component and the
asymptot) have their own physical origin. The asymptot
represents a minimum value for the UV slope $\beta$ and IRX. The
minimum value for $\beta$ is set by the
stellar population itself and corresponds to stellar population age. The minimum value
for IRX equals zero, corresponding to a scenario where no
emission is absorbed.  

The physical origin of the power law can be understood when focusing
on the absorption of stellar emission by dust. For a uniform screen of
dust the nominator and denominator
of IRX  = $L_{\rm IR}/L_{1600 \Angstrom}$
scale exponentially with the optical depth of the
screen, such that 
\begin{equation*}L_{\rm IR}\propto
\int_{\lambda,\rm{min}}^{\lambda,\rm{max}} [1 - \exp{(-\tau_{\lambda})}]
d\lambda
\end{equation*}
and 
\begin{equation*}
L_{\rm 1600 \Angstrom}\propto \exp{(-\tau_{1600\Angstrom})}.
\end{equation*}
The UV
slope $\beta$ is set by the slope between FUV and NUV luminosity. Both of these scale exponentially as
\begin{equation*}
L_{\lambda,\rm{NUV|FUV}} \propto
\exp{(-\tau_{\lambda,\rm{NUV|FUV}})}.
\end{equation*}
If we express the optical depths
in terms of the V-band optical depth ($\tau_{\lambda} = A(\lambda) \times
\tau_{\rm V}$, where $A(\lambda)$ is a function of wavelength),
it immediately becomes clear that both IRX and $\beta$ exponentially
scale with the V-band optical depth $\tau_{\rm V}$. This makes the
exponential dependence on $\tau_{\rm V}$ the
driver of the
power law trend. In this framework it also becomes clear that as the
ratio $A_{\lambda,\rm FUV}/A_{\lambda,\rm NUV}$ increases, the slope
of the relation between UV
slope $\beta$ and $\tau_{\rm V}$ steepens. This is the principal cause
of the flattening in the \IRXB relation for dust attenuation curves with an increased
FUV/NUV attenuation ratio (e.g., an SMC curve).


\subsection{Observed variations in the \IRXB relation}
\subsubsection{A spatially resolved \IRXB relation for local galaxies}
\citet{Boquien2012} looked at the \IRXB relation on sub-galactic
scales in star-forming galaxies as a part of the \emph{Herschel}
reference survey \citep{Boselli2010}. The comparison between the
Boquien et al. results and our model is interesting, as sub-galactic
regions are perhaps more similar to the screen of dust experiment
presented in this work than integrated galaxy
properties. \citet{Boquien2012} find that for all except one galaxy in
their sample, the \IRXB relation on sub-galactic scales is located
below the \citet{Meurer1999} and \citet{Overzier2011} relation for
local (blue starburst) galaxies. Our results suggest that stellar
age and differences in the dust attenuation curves may cause such variations. Only one galaxy in their sample, NGC 4536
which is undergoing a nuclear starburst, follows
the relation for starburst galaxies. An increased level of turbulence
or a two-component dust model in starburst environments would naturally cause such a shift, as would
a young light-weighted mean stellar age of a few Myr.

Boquien et al. fit SEDs to study variations in the \IRXB relation as a function
of a number of parameters, including stellar mass, SFR, 4000 \Angstrom
break index, V-band attenuation, FUV attenuation, mass weighted
age, a parameter $\alpha$ tracing dust temperature \citep{Dale2002},
and the slope of the attenuation curve $\delta$.  The authors found
that deviations from the \citet{Meurer1999} \IRXB relation are poorly
described by age sensitive parameters such as the 4000 \Angstrom break
index, birthrate parameter, and the mass normalised age. \citet{Boquien2012}
find that variations can best be described by variations in the intrinsic UV colour in the absence
of dust between different sub-galactic regions. Such variations
originate from differences in star-formation history (SFH) of the
respective regions. \citet{Boquien2012} furthermore conclude that the shape of the
attenuation curve can play a secondary role in describing deviations
from the \citet{Meurer1999} relation. The latter is in agreement with
our results of a flattening in the \IRXB relation as the FUV/NUV
attenuation ratio increases. The former, intrinsic variations in
$\beta$ due to differences in SFH, is not explored in our
model. Such variations should reflect themselves in
variations in the 
light-weighted age (rather than mass normalised age). The light-weighted stellar age can
still be a key player that causes the redder UV colours for sub-galactic
regions.

\citet{Munoz-Mateos2009} looked at the \IRXB relation in radial bins
of local galaxies from the \emph{Spitzer} Infrared Nearby Galaxies
Survey sample \citep{Kennicutt2003} and found that both $\beta$ and
IRX correlate well with gas-phase metallicity. Observations and
theoretical models have demonstrated a clear relation between
gas-phase metallicity and dust-to-gas ratio
\citep[e.g.,][]{Leroy2011,Remy-Ruyer2014,Popping2016}. An increase in
gas-phase metallicity therefore corresponds to an increase in the
dust-column and consequently the optical depth seen by the
photons. Since both $\beta$ and IRX scale exponentially with V-band
optical depth, this
naturally results in a trend between $\beta$ and IRX and gas-phase
metallicity (as long as the gas column the emission travels through
doesn't change dramatically from one region to the other). Like
\citet{Boquien2012}, \citet{Munoz-Mateos2009} find that starburst
galaxies have bluer UV colours than normal star-forming galaxies. As
stated before, turbulence driven star formation, a two-component dust
model, and young stellar ages naturally cause such blue UV colours.

\citet{Ye2016} looked at the \IRXB relation for individual \hii
regions in NGC 628. Individual \hii regions may be the fairest
observational comparison to our model of a screen of dust in front of a single stellar
population. \citet{Ye2016} find that H$\alpha$ equivalent width (an
age tracer) significantly correlates with deviations from the local
(blue starburst) \IRXB relation. As the equivalent width decreases
(older stellar populations) the \hii regions have redder UV colours
$\beta$. Ye et al. furthermore find that stellar
population age and the 4000 \Angstrom break also moderately correlate
with deviations. This is in good agreement with our conclusion that
older stellar populations drive variations towards red UV colours \citep[see also,][and
many others]{Kong2004,Grasha2013}. \citet{Ye2016} also find that
metallicity and Balmer decrement show no correlation with deviations
away from the \IRXB relation.

\subsubsection{SMC dust in early galaxies}
\citet{Capak2015}, \citet{Bouwens2016}, \citet{Pope2017}, and \citet{Smit2017} observed a number of high-redshift ($z>4$)
galaxies with a UV-slope $\beta \sim -1$ and IRX$\sim1$ or less. These
works concluded that these values are consistent with an SMC
attenuation curve, suggesting that dust in early Universe
galaxies is similar to SMC dust. We found that deviations towards red UV-slopes with respect to the
canonical \IRXB relations are driven by either
an ageing stellar population or by varying dust types. Similar values for $\beta$ and IRX
as observed by \citet{Capak2015}, \citet{Bouwens2016}, 
\citet{Pope2017}, and \citet{Smit2017}, can be obtained when the stellar emission from a 50--100 Myr old stellar population passes through
a screen of dust with a MW attenuation curve. The effect of
ageing has been discussed before by many authors \citep[e.g.,][]{Kong2004,Grasha2013}.

Without additional information on the age of the observed stellar population
it is hard to disentangle dust attenuation curve versus
age. Some additional insight can still be gained when taking the
typical lifetime of molecular clouds into account. Assuming
that stars leave their birth clouds after 5--20 Myr \citep[the
lifetime of molecular clouds,][]{Murray2010}, strong deviations
towards red UV-slopes of $\beta=-1$ for low values of IRX 
are more likely to be driven by changes in the dust attenuation curves
(the stellar populations are too young to cause significant shifts in
the \IRXB relation). {Furthermore, older stellar
  populations should not be a large concern at higher redshifts given
  the much smaller stellar population ages and much higher specific
  star formation rates in galaxies at high redshifts.} The bright UV and IR
emission (and in some cases $\rm{[CII]}$ emission) detected in early
Universe galaxies indeed suggest that these galaxies are actively star
forming and likely have a UV spectrum dominated by young stars. This
favours an SMC attenuation curve over a MW curve for the $z>4$
galaxies.

\citet{Narayanan2017} suggested that the low IRX values
observed in $z>4$ galaxies are due to a poor assumption for
the dust temperature, rather than SMC dust attenuation curves \citep[see
also][]{Ferrara2016}. When only a handful of photometric datapoints
are available for the IR wavelength regime, the total infrared
luminosity is estimated based on an assumption for the dust
temperature. \citet{Narayanan2017} compared the dust temperatures of galaxies in hydrodynamical models
to assumed dust temperatures and found that the assumed temperatures
for high-redshift galaxies are often lower than the dust temperatures suggested by the models. An increase in the assumed dust temperature naturally increases the total
infrared luminosity and IRX towards a MW \IRXB relation. \citet{Bouwens2016} also concludes that the
dust temperature of galaxies might increase with look-back time, based on
the large number of non-detected galaxies at 3mm continuum in ASPECS
\citep[The ALMA Spectroscopic Survey in the Hubble Ultra Deep
Field,][]{Walter2016}. A better photometric sampling of the
Rayleigh-Jeans tail of the sub-mm SED of high-redshift galaxies is
necessary and will
provide valuable insights. 

Recently, \citet[see also \citet{Reddy2012}]{Reddy2017} found that
normal star-forming galaxies at $1.5\leq z \leq 2.5$ have values for IRX and
$\beta$ that correspond best to an SMC dust attenuation
curve. \citet{Siana2008} and \citet{Siana2009} found similar results
for lensed Lyman Break galaxies at $z\sim3$. The hypothesis of  a dust temperature that is too low can not be valid for the
SMC attenuation curves favoured in these works. Good \emph{Spitzer} and \emph{Herschel} coverage of
the peak of the IR SEDs is available for these
galaxies. A lack of information about the SED at
millimetre wavelengths can still introduce a bias towards low IRX
values. The \emph{Herschel} photometry is sensitive to the warm dust,
and may miss the contribution to the total IR luminosity from a large
cold dust component (although more than half of the IR flux has to
come from the cold component for IRX to align with a MW dust
curve). 

There appears to be some tension between the
 Siana et al. and Reddy et al. results and the result of studies of the \IRXB relation for bright UV-selected
galaxies at the same redshifts that found an \IRXB relation similar to
the local (blue starburst) relation \citep[e.g.,][]{Heinis2013,
  Coppin2015,Alvarez-Marquez2016}. Differences in SFH may
play a role here (for instance young stars driving the bright UV emission in
UV-selected galaxies), but there
is evidence that differences in the attenuation curves may really be the key driver of
the SMC \IRXB relation. Kriek \& Conroy fit composite SEDs with flexible
stellar population synthesis models to galaxies, while exploring attenuation
curves with varying slopes and UV bump strengths \citep[see also][]{Salmon2016}. They find
that star-forming galaxies at $z\sim2$ on
average have an attenuation curve with a slope between MW and SMC and
a weak presence of a 2175 \Angstrom bump (25\% of the MW
bump). They furthermore find that galaxies with larger specific SFRs
have shallower extinction curves. Unfortunately our model is not
equipped to relate different dust types to galaxy properties such as
stellar mass and SFR. 

If any conclusion has to be reached based on the apparent SMC \IRXB
relation in $z>4$ galaxies and star-forming galaxies at $z\sim2$, it
is that good coverage of the entire SED of galaxies from the UV to the sub-mm is necessary to properly asses the \IRXB relation in
different types of galaxies. The good coverage allows for a careful
determination of both $\beta$ and IRX, while simultaneously providing
constraints on the dust attenuation curve. 

If dust attenuation curves vary between SMC and MW type from one
galaxy to the other, this is an interesting result in itself. The SMC
dust model is generally thought to better represent the dust properties of galaxies
in the early Universe than a MW dust model. This is motivated by the low metallicities these galaxies are thought to
have and a dust population dominated by dust produced in SNe and
through the accretion of metals onto dust grains \citep[e.g.,][]{Zhukovska2013,Popping2016}. Massive main-sequence galaxies at $z\sim2$ on the other hand
have metallicites close to solar \citep[e.g.,][]{Zahid2013} and
therefore this
argument doesn't hold for these objects. Furthermore, at these
redshifts dust produced in active giant branch stars becomes
increasingly more important, likely changing the dust attenuation curve. Shattering
of grains promotes the production
of small grains and strong FUV attenuation \citep{Hirashita2010}.
This may play a key role in the turbulent ISM of massive and actively star-forming
galaxies at $z\sim2$ and thus be the real driver of SMC attenuation
curves and SMC \IRXB relations. At the same time shattering can
be balanced by the coagulation of grains in dense
media \citep{Hirashita2014,
  Hirashita2015} and result in MW type attenuation curves. Besides
  observational efforts, it is thus of key
  importance that cosmological galaxy formation models address the origin of dust attenuation curves in
different galaxy types over cosmic time, accounting for the many processes that
shape attenuation curves (e.g., different dust formation mechanisms
and grain-grain collisions).

\begin{table*}
 \caption{\label{tab:fit_params}Best fit parameters for the analytic
   approximation of the \IRXB relation invoking dust type, stellar population age, and
   turbulence of the dust screen (Equations \ref{eq:fit1} to \ref{eq:fit2}).}
 \begin{tabular}{lcccccccccccccccc}
\hline
\hline
Dust type && $A_{\rm screen}$ & $B_{\rm screen}$ & $C_{\rm screen}$&&&
                                                                    $A_{\rm
                                                                    turb}$&$B_{\rm
                                                                            turb}$&
                                                                                    $C_{\rm
                                                                                    turb}$&
                                                                                            $D_{\rm
                                                                                            turb}$  &  $E_{\rm
                                                                                            turb}$  &&&$A_{\delta\beta}$ &$B_{\delta\beta}$ & $C_{\delta\beta}$\\
      
\hline
MW && 0.69 & 4.17 & 1.72 &&& 0.05 & 5.89 & 2.8 & 0.41 & 0.37 &&& 0.05 & 0.74 & 1.11 \\
MW-BUMP00 && 0.45 & 4.65 & 1.89 &&& 0.03 & 5.44 & 2.52 & 0.44 & 0.41 &&& 0.08 & 0.72 & 1.14 \\
LMC && 0.78 & 2.66 & 1.09 &&& 0.06 & 3.53 & 1.76 & 0.42 & 0.35 &&& 0.17 & 0.7 & 1.24 \\
SMC && 0.66 & 1.84 & 0.75 &&& 2.01 & -0.06 & 0.64 & -2.24 & 0.66 &&& 0.39 & 0.63 & 1.54 \\                                                            
\hline
 \end{tabular}
 \end{table*}

\subsubsection{Geometry effects make galaxies blue}
\label{Sec:disc_blue}
Observations of high-redshift dusty star-forming galaxies have
suggested that galaxies become bluer as their infrared luminosity
increases \citep[e.g.,][]{Penner2012,Oteo2013,Casey2014,
  Bourne2017}. In our model there are multiple competing effects that
can cause the UV-slope $\beta$ to shift to lower values at fixed IRX
(turbulence, a small fraction of stellar emission not covered by dust,
a uniform mixing of stars within the dust screen, two-component dust
model). For example, a two-component dust model can describe the location in the
\IRXB plane of a large fraction of the blue low- and high-redshift
DSFGs. Given that DSFGs are very dust-rich systems, the effects of the
two-component dust model and the \IRXB relation are very likely to
play an important role (a large optical depth of the `diffuse' dusty ISM).

Although with
varying degree, all these aforementioned effects result in some fraction of the
stellar emission being completely absorbed and driving IRX, whereas
the remaining fraction of the original stellar emission is hardly absorbed and responsible for the
observed UV emission.  For young stellar populations, this results in
blue UV-slopes $\beta$. In summary, blue galaxies with a strong IR
component are the result of spatial variations in the optical depth
through which the stellar emission travels \citep[see also
e.g.,][]{Casey2014}. \citet{Koprowski2016} and \citet{Chen2015} have
shown significant offsets between the origin of the rest-UV/optical
and sub-mm emission in DSFGs. Similarly,
  \citet{Puglisi2017} showed that in z $\sim$ 1.6 off-main-sequence
  galaxies the Balmer decrement (i.e. an optical dust attenuation
  tracer) yields a dust attenuation much lower than the attenuation one
  would derive from the observed IRX. These results demonstrate that UV/optical and IR
emission can originate from very different regions and therefore can
not be approximated by a single uniform screen of dust. Similar conclusions were
reached by other theoretical efforts to understand the nature of
blue DSFGs \citep{Safarzadeh2017,Narayanan2017}, although these works
did not explore all the geometry effects presented in this work.

It is worthwhile to focus a bit more on the effect of turbulence on the
\IRXB relation. An increased level of turbulence may be important for dust absorption
in stellar birth clouds. Giant Molecular Clouds (GMCs) in our Milky
Way and in local galaxies have significant non-thermal line widths
\citep[e.g.,][]{Fukui2001, Elmegreen2004}, a signature of the presence of supersonic
turbulence. \citet{Swinbank2011} found a highly
turbulent ISM in a dusty
star-forming galaxy at $z=2$. Supersonic turbulence in the birth clouds of young stars
can thus naturally cause deviations away from the canonical \IRXB relation and
can be responsible for at least a fraction of the scatter around the
observed \IRXB relations. \citet{Casey2014} found that
local galaxies with SFRs larger than 25 $\text{M}_\odot\,\text{yr}^{-1}$
are typically located above their observed best-fit \IRXB relation for
local galaxies (i.e., bluer than expected) and that the deviation from
the canonical \IRXB relations for high redshift DSFGs increases as
a function of FIR luminosity. Empirical fits of the \IRXB relation for local
starbursts \citep{Meurer1999,Overzier2011} suggest bluer UV colours
than fits to local 'normal' star-forming galaxies \citep[see also
\citet{Boquien2012}]{Casey2014}.  An increased level of turbulence in
the birth clouds or the diffuse ISM of starbursts is one of the
possible explanations for this.

Unfortunately, without detailed
measurements of the turbulence within molecular clouds and the diffuse ISM it is hard to
quantify how much of the deviations away from the \IRXB relation can be accounted for by a turbulent
dust screen. Absorption studies probing non-thermal line widths towards turbulent regions in the
 Milky Way or local galaxies could potentially shed more light on
 this. The high-spatial resolution
capabilities of the Atacama Large (sub-)Millimeter Array (ALMA) can
resolve GMCs in nearby galaxies. This makes ALMA an ideal instrument
for such an exercise. An interesting alternative would be to study deviations from the \IRXB
relation in galaxies as a function of star-formation efficiency. The turbulence within molecular clouds is intimately
related to the efficiency with which stars form
\citep{Krumholz2005}. One could thus expect galaxies to become bluer
as the star-formation efficiency increases (or the molecular
hydrogen depletion time decreases). It is furthermore worthwhile to study deviations from the
 \IRXB relation as a function of galaxy velocity dispersion.

We want to spend a few more words focusing on the two-component dust
model, in which dusty birth clouds are homogeneously distributed within a
diffuse dusty ISM. First of all, the effects of this model only become
important once the optical depth $<\tau_{\rm V}>$ of the diffuse dusty ISM is larger
than approximately four. As stated before, this limits the importance
of this model to very dust-rich systems (e.g., DSFGs). Locally, reddening of light
in birth clouds is approximately 2 times stronger than reddening of
light in the diffuse ISM \citep{Calzetti2000}. \citet{Puglisi2016}
showed that for far-IR selected normal star-forming galaxies at $z\sim1$ the reddening of light in birth clouds is
roughly similar to the reddening of light in the diffuse ISM. This
suggests that birth clouds have a larger filling factor in galaxies at
$z\sim1$ than locally.

\begin{figure*}
\includegraphics[width = 0.65\hsize]{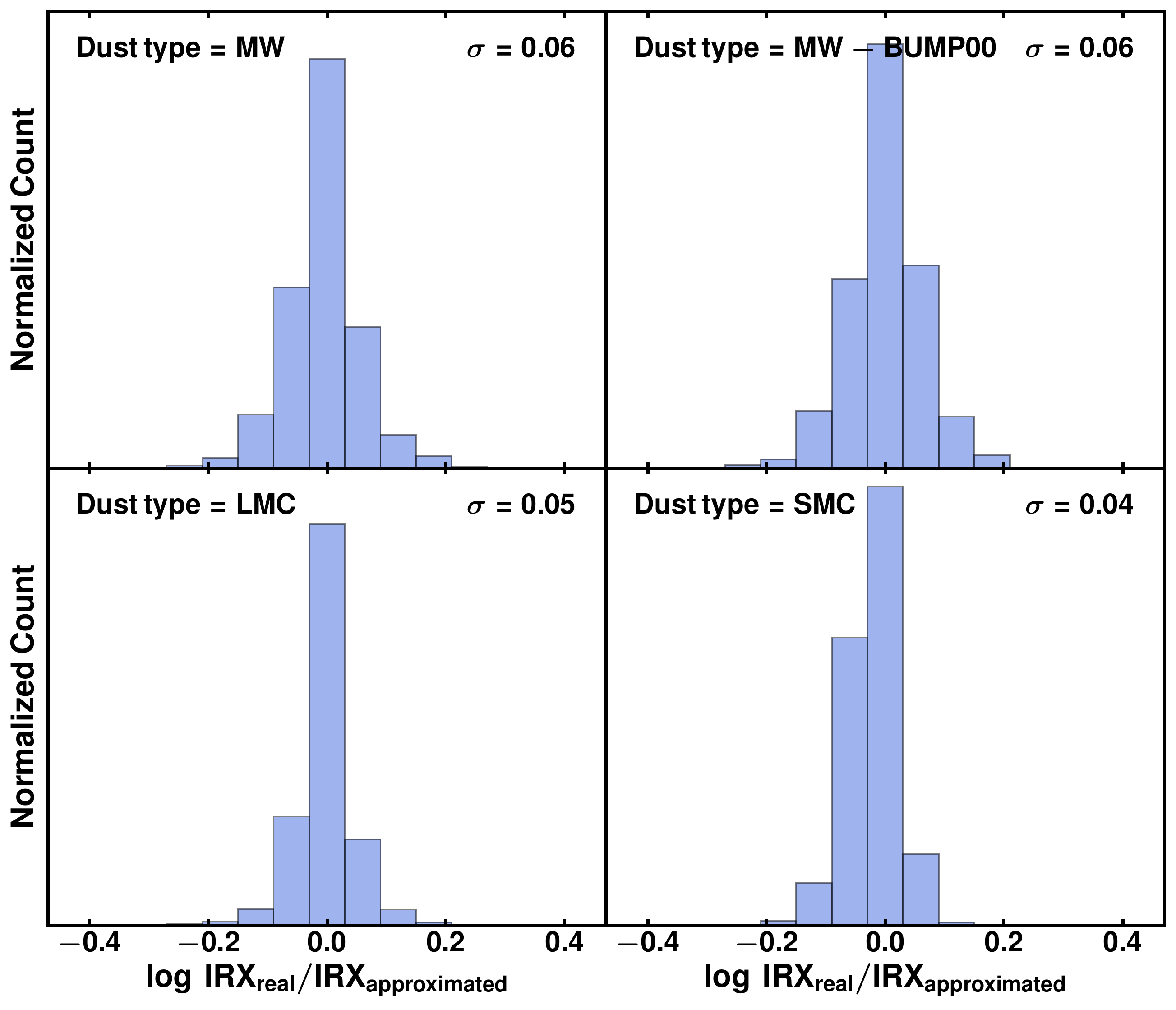}
\caption{The distribution of the log
of the uncertainty (i.e. $\log [\rm{IRX}_{\rm real}/\rm{IRX}_{\rm
  approximated}]$)  in IRX based on the analytic approximation for the \IRXB relation
(Equations \ref{eq:fit1} to \ref{eq:fit2}) for the four different
dust types discussed in this work. This plot was created by
running 25 000 combinations of homogeneous V-band optical depth
($<\tau_{\rm V}>$
running from 1 to 20),
Mach number (running from $\mathcal{M} = 0$ to $\mathcal{M} =4$), and
stellar population age (from 5 to 100 Myr) for each dust type (see
Sec. \ref{sec:IRXB_fit} for a detailed description). 
\label{fig:fit_error}}
\end{figure*}

\subsection{Analytic approximation for the \IRXB relation}
\label{sec:IRXB_fit}
One of the main motivations for the \IRXB relation is that it can
potentially account for absorbed UV emission. Variations in the \IRXB relation unfortunately complicate this. Only if these deviations
can be described as a function of other parameters can the \IRXB
relation really be used to account for obscured SF.

\citet{Safarzadeh2017} conclude that
variations in the dust composition are the dominant source of variations in the \IRXB relation in their
simulations. \citet{Kong2004} suggested that the ratio of the present to past
average SFR defines the deviation of galaxies away from the canonical
Meurer et al. \IRXB
relation. This was contradicted by \citet{Johnson2007},
\citet{Boquien2012}, and \citet{Johnson2013} who find
that the dispersion in UV-slope $\beta$ hardly correlates with the
star-formation history of galaxies as traced through their 4000
\Angstrom break. \citet{Grasha2013} found that deviations away from
the \IRXB relation can be described by the mean stellar age of
the galaxies \citep[but see][]{Boquien2012}. We also find that the stellar population age drives the
UV-slope $\beta$ towards higher values. 

\citet{Narayanan2017} concludes that besides age, the total SFR (or IR luminosity) of a galaxy is an
important parameter that controls the deviation of galaxies away from
the canonical \IRXB relation. This would be in line with the $z=0$ sample
presented in \citet{Casey2014}, who found that galaxies lie
increasingly above the best-fit \IRXB relation as a function of SFR
and IR luminosity. As discussed in Section \ref{Sec:disc_blue},
the underlying physical motivation for this might be an increase in
the turbulence within the birth clouds of young stars.

Our results show that the origin of deviations away
from local fits of the \IRXB relation is much more complicated and
can not simply be captured by a combination of stellar age and
SFR (turbulence). Besides these two parameters, the attenuation curve of the
screen of dust, the mixing of stars within the screen of dust, a
two-component dust model, and the
covering fraction of the dust all contribute heavily to deviations
away from the \IRXB relation. As an experiment we could imagine a
galaxy for which a UV-slope $\beta$ of -1 is observed and a mean
stellar age of 10 Myr is determined. If only turbulence would play a role,
the range in IRX one could get covers approximately 0.2 dex for a MW
dust attenuation curve. The assumption of a two-component dust model
or that stars and dust are mixed increases the range in
acceptable IRX values to about $\sim 0.6$ dex. If no information is
available for the dust attenuation curve of the galaxy, the range in
possible IRX values increases to $\sim 1.5$ dex. Without prior
information on the covering fraction of the dust the range of possible
IRX values increases up to three orders of magnitude. Such variations
are not covered anymore by two parameters, but require detailed
spatially resolved information about the galaxy and full SED fitting
to constrain the most likely dust attenuation curve
\citep[e.g.,][]{Noll2009,Kriek2013, Salmon2016}.

An additional complication to the parametrisation of deviations are
observational effects. As demonstrated in Section
\ref{sec:observational_effects}, the  sampling of the
stellar SED by telescope filters can lead to systematic variations in
the location of galaxies within the \IRXB plane. Lastly, ideally
deviations away from the \IRXB relation can be parametrised as a
function of properties that can be determined from the UV spectrum
alone, unlike age and total SFR. 

Despite the negative tone, it is still worthwhile to find an approximate analytic equation for the \IRXB relation that
includes a subset of the discussed physical processes. Here we focus
on an approximation that includes the effects of stellar
population age and turbulence for  the four different attenuation
curves discussed in this work. These approximations can be useful in case good arguments can be made that a screen of dust
without holes is a good approximation of the dust distribution in a
galaxy (e.g., if the non-obscured UV emission escaping through these holes is too faint
to observe). Accounting for stellar age and turbulence we can fit the relation between IRX and $\beta$
using the following set of equations:
\begin{multline}
\label{eq:fit1}
\rm{IRX}(\beta,\rm{Age},\mathcal{M})=\\ \rm{IRX}_{\rm screen}(\beta,\rm{Age}) + \rm{IRX}_{\rm turb}(\beta,\rm{Age},\mathcal{M})
\end{multline}
where
\begin{multline}
\rm{IRX}_{\rm screen}(\beta,\rm{Age})   = \\A_{\rm screen} \times \biggl[10^{0.4 [B_{\rm
    screen} + C_{\rm screen}(\beta - \delta\beta)]} - 1\biggr],
\end{multline}
\begin{multline}
\rm{IRX}_{\rm turb}(\beta,\rm{Age},\mathcal{M}) = \\A_{\rm turb} \times \biggl[ 10^{0.4[B_{\rm
    turb}\mathcal{M}^{D_{\rm turb}} + C_{\rm turb}\mathcal{M}^{E_{\rm
      turb}}(\beta - \delta\beta)]}-1\biggr],
\end{multline}
and 
\begin{equation}
\label{eq:fit2}
\delta\beta({\rm Age} )= \exp{[A_{\delta\beta} + B_{\delta\beta}\times
  \log{(\rm{Age/10\,Myr})}]} - C_{\delta\beta}.
\end{equation}
We set $\rm{IRX}_{\rm turb}$ to zero when $\rm{IRX}_{\rm turb} <
0$ (as this marks an unphysical scenario with negative IR luminosities). IRX$_{\rm screen}$ marks the IRX one would get for a uniform screen
of dust. IRX$_{\rm turb}$ is a correction that accounts for the
turbulence within the screen. When $\mathcal{M}=0$, a uniform screen, IRX$_{\rm turb}$
will automatically equal zero. $\delta\beta({\rm Age})$ takes the shift in
$\beta$ due to stellar population age into account. The best-fit
parameters for Equations \ref{eq:fit1} to \ref{eq:fit2} are listed in Table \ref{tab:fit_params}.

We tested our analytic approximation for 25 000 different
combinations of homogeneous V-band optical depth (
running from 1 to 20),
Mach number (running from $\mathcal{M} = 0$ to $\mathcal{M} =4$), and
stellar population age (from 5 to 100 Myr).  We first
  calculated the real IRX and $\beta$ for each of these combinations using the approach outlined in Section \ref{sec:model}. This resulted in IRX$_{\rm
    real}$ and $\beta_{\rm real}$. We then used the analytic
  approximation presented in Equations \ref{eq:fit1} to \ref{eq:fit2}
  to calculate IRX$_{\rm approximated}$ as a function of $\beta_{\rm
    real}$, as well as the stellar population age and Mach number that
  went into the calculation for IRX$_{\rm real}$.

We quantify the uncertainty in IRX$_{\rm approximated}$ in Figure
\ref{fig:fit_error} by showing the difference between the real and approximated IRX ($\log [\rm{IRX}_{\rm real}/\rm{IRX}_{\rm
  approximated}]$) for a given
$<\tau_{\rm V}>$, stellar population age, and Mach number (and $\beta_{\rm real}$). The typical
one-sigma uncertainty in IRX$_{\rm approximated}$ is $\sim$ 0.05 dex,
which suggests that our analytic expression is a good
representation of the \IRXB relation. If turbulence is
indeed correlated with SFR, non-thermal line width, SFE, or velocity
dispersion, similar empirical equations could be derived as a
function of these galaxy properties.  We remind the reader that the presented
analytic equation does not include the effects of a patchy dust screen
or the mixing of dust and stars.

\subsection{Future prospects to use the \IRXB relation to estimate the
  intrinsic SFR of galaxies}
Although the large theoretical scatter in the \IRXB relation seems to
be discouraging for its use as a tracer of absorbed UV emission,
this does not necessarily need to be. If the \IRXB
relation for galaxies of certain types at certain epochs can be constrained, these
constrained relations may still be valuable when estimating the
obscured UV emission of galaxies. For instance, the IR-brightest $z\sim2$ DSFGs
appear to suffer more from dust geometry effects than normal
star-forming galaxies at these redshifts
\citep[e.g.,][]{Penner2012,Oteo2013,Casey2014}. High spatial
resolution observations and/or good SED coverage of the rest-frame UV
wavelength over large redshift ranges make  detailed studies of the \IRXB relation for
individual galaxy populations over cosmic time possible.

NIRCam on board of
the \emph{James Webb Space
  Telescope} (JWST) will
offer up to 0.1 arcsec imaging of the entire UV rest-frame wavelength
regime of galaxies at $z\geq3$. Combined with the exquisite spatial resolution of ALMA to probe the sub-mm emission of galaxies, this opens up
the possibility of a detailed resolved study of large scale dust geometry effects  in galaxies (differences in
location of the UV and IR emission). The good photometric sampling of the
rest-frame UV of galaxies at $z\geq3$ by NIRCam enables SED fitting of the rest-frame
UV spectrum. This provides essential insights into the properties of the
dust (e.g., the presence of a 2175 \Angstrom bump and the slope of the attenuation
curve).  Large scale geometry effects and differences in the dust attenuation curve
are among the physical processes that cause largest variations in the
\IRXB relation. The combination of JWST and ALMA to obtain spatially
resolved imaging is thus a perfect approach to constrain the \IRXB relation for different types of galaxies over a large
redshift range (including normal star-forming galaxies, DSFGs, and
galaxies during the epoch of reionization), where different physical processes may dominate. 

Other instruments such as the Euclid and the Wide-Field Infrared Survey
Telescope (WFIRST) will provide good sampling of the
rest-frame UV of high-redshift galaxies on large areas on the sky (though at a lower spatial
resolution than JWST and depending on the redshift the rest-frame UV spectrum is not entirely covered). Unfortunately, ALMA is not in the
position to follow up all the sources that will be detected by these
future NIR space observatories (including JWST).  Nevertheless,
constraints on the dust attenuation curves in different classes of
galaxies through the SED fitting of the photometric sampling of the
rest-frame UV on itself is valuable to better constrain the \IRXB
relation of galaxies (and in general the dust properties).

Although challenging, future small sample studies to constrain the \IRXB
relation and/or some of the physical processes that cause variations for different galaxy
types at different epochs will justify the use of the \IRXB relation
to reliably estimate the obscured UV emission and intrinsic SFR of galaxies.

\section{Summary \& Conclusions}
\label{sec:summary}
The \IRXB relation is commonly used to
estimate the intrinsic SFR in galaxies when no IR/sub-mm
information is available. In this paper we have presented a simple model to explore how
different physical effects cause variations in the \IRXB dust attenuation relation. We place a screen of dust in between
a young stellar population and the observer and vary the properties of
the dust screen and the age of the stars. The attenuation of stellar light
is modelled by applying attenuation curves presented in
\citet{Seon2016}, who perform radiative transfer calculations in a
spherical, clumpy interstellar medium. We summarise our main findings below:
\begin{itemize}
\item Our simple model can reproduce the observed \IRXB relation for local (blue starburst) galaxies as presented in \citet{Meurer1999},
  \citet{Overzier2011}, and \citet{Casey2014} when invoking a uniform
  screen of MW type dust in front of a stellar population with an age
  of $\sim$10 Myr. The asymptotic power law relation between IRX and
  $\beta$ as found in the literature and suggested by our models is
  driven by the exponential dependence of both $\beta$ and IRX on
  the V-band optical depth.
\item An increase in stellar population age increases the UV slope
  $\beta$, and moves galaxies to higher values of $\beta$ for
  a fixed IRX.
\item An increase in the FUV absorption with respect to longer
  wavelengths flattens the \IRXB dust attenuation relation, driven by a
  steeper UV slope $\beta$. This
  effect is apparent when comparing the \IRXB relation for MW, LMC,
  and SMC dust types.
\item An increased level of turbulence in the screen of dust leads to
  lower values for both $\beta$ and IRX, in such a way that galaxies
  move towards locations in the \IRXB plane above the relation for a
  uniform dust screen. For a fixed IRX, galaxies become bluer.
\item A patchy distribution of dust with holes in it causes strong
  deviations from the \IRXB relation for a uniform dust screen, with
  blue UV colours combined with high values for IRX.
\item When stars are mixed in between the screen of dust
  rather than placed in front of it, both $\beta$ and IRX decrease,
  causing deviations towards bluer UV colours for a given IRX. 
\item A two-component dust model, where dusty birth clouds are
  distributed within the diffuse ISM, also causes deviations towards
  blue UV colours for a given IRX.
\item A poor photometric sampling of the rest-frame UV spectrum can
  artificially move galaxies around the \IRXB plane. A remedy for this
  is a filter combination that at least probes the rest-frame FUV
  ($\sim 1250$ \Angstrom) and rest-frame NUV ($\sim 3000$ \Angstrom) wavelengths.
\end{itemize}
Taken together, dust geometry effects cause variations towards blue UV
colours, whereas strong FUV absorption and stellar age
cause variations towards red UV colours for a fixed IRX.  

Our results demonstrate that there are numerous competing processes
that may simultaneously take place that can move galaxies around the \IRXB
plane. More simply put, there exists no universal \IRXB dust
attenuation relation. We present an analytical approximation for
the \IRXB relation when there are good reasons to believe that all the
stellar emission is (up to some extent) absorbed by dust. High-resolution imaging studies of
the \IRXB relation combining for instance NIRCam on board
of JWST with ALMA will constraint the \IRXB relation for different
galaxy types at different epochs where different physical processes that
cause variations dominate. These studies are challenging,
but in turn will open up the reliable use
of the \IRXB relation to estimate the intrinsic SFR of galaxies,
despite the lack of a universal \IRXB dust attenuation relation.

\section*{Acknowledgements}
We thank the anonymous referee for useful comments that helped
clarify this paper. We thank Chian-Chou Chen, Richard Ellis, Katharina Immer, Desika Narayanan, Ralf
Siebenmorgen, Wuji Wang, and Anita Zanella for valuable conversations. We thank
Desika Narayanan and especially Richard Ellis for commenting on an earlier version of this paper. GP thanks Claus
Leitherer for his support in running \texttt{Starburst99}. CN thanks ESO for its
hospitality during a long-term visit.
\bibliographystyle{mn2e_fix}
\bibliography{references}

\appendix
\section[]{Deviations from the \IRXB relation due to observational
  effects: Additional filter combinations}

\begin{table*}
 \caption{\label{tab:filter_combs} A suggestion for
   combinations of filters (and imaging instruments) for
   the rest-frame FUV and NUV wavelength range that result in a reliable
   reconstruction of the UV slope $\beta$ and the \IRXB
   relation. Any additional filters with central wavelengths in the
   rest-frame UV will improve the quality of the reconstruction. Other
 combinations are of course possible.}
 \begin{tabular}{ cp{8cm}p{8cm}}
\hline
\hline
 redshift & rest-frame FUV wavelengths& rest-frame NUV wavelengths\\
\hline
\vspace{2mm}
0.0&\Galex FUV & HST UVIS-F275W\\
\vspace{2mm}
0.5&\Galex NUV & HST ACS-F475W, Subaru B-band, g-band\\
\vspace{2mm}
1.0&HST UVIS-F275W & HST ACS-F606W, Subaru r-band\\
\vspace{2mm}
1.5&HST UVIS-F336W, Subaru U-band & Subaru i-band\\
\vspace{2mm}
2.0&HST ACS-F475W, Subaru B-band, g-band & HST ACS-F814W,
                                                        Subaru z-band,
                                           NIRCam\\
\vspace{2mm}
2.5&HST ACS-F475W, Subaru B-band, g-band & HST
                                                        WFC3-F105W,
                                                        MUSE, NIRCam\\
\vspace{2mm}
3.0&HST ACS-F606W, Subaru V-band, NIRCam & HST WFC3-F125W, NIRCam\\
\vspace{2mm}
4.0&HST ACS-F775W,Subaru r-band, NIRCam & HST
                                                                WFC3-140W,  WFC3-160W, NIRCam\\
\vspace{2mm}
5.0&HST ACS-F814W, Subaru i-band, NIRCam & H-band, NIRCam\\
\vspace{2mm}
6.0&HST WFC3-105W, Subaru z-band, NIRCam & Ks-band, NIRCam\\
\vspace{2mm}
7.0&HST WFC3-105W, NIRCam& Ks-band, NIRCam\\
\vspace{2mm}
8.0&HST WFC3-125W, NIRCam & NIRCam\\
\hline
 \end{tabular}
 \end{table*}
\label{sec:appendix}
We explored the effect of poor photometric sampling of the rest-frame
UV spectra on the derived \IRXB relation in Section
\ref{sec:observational_effects}. The exact choice of filters affects the quality of the sampling of the rest-frame UV spectrum. Here
we present the sampling of the rest-frame UV spectrum and the
resulting \IRXB relation using different combinations of filters.

\begin{figure*}
\includegraphics[width = 0.8\hsize]{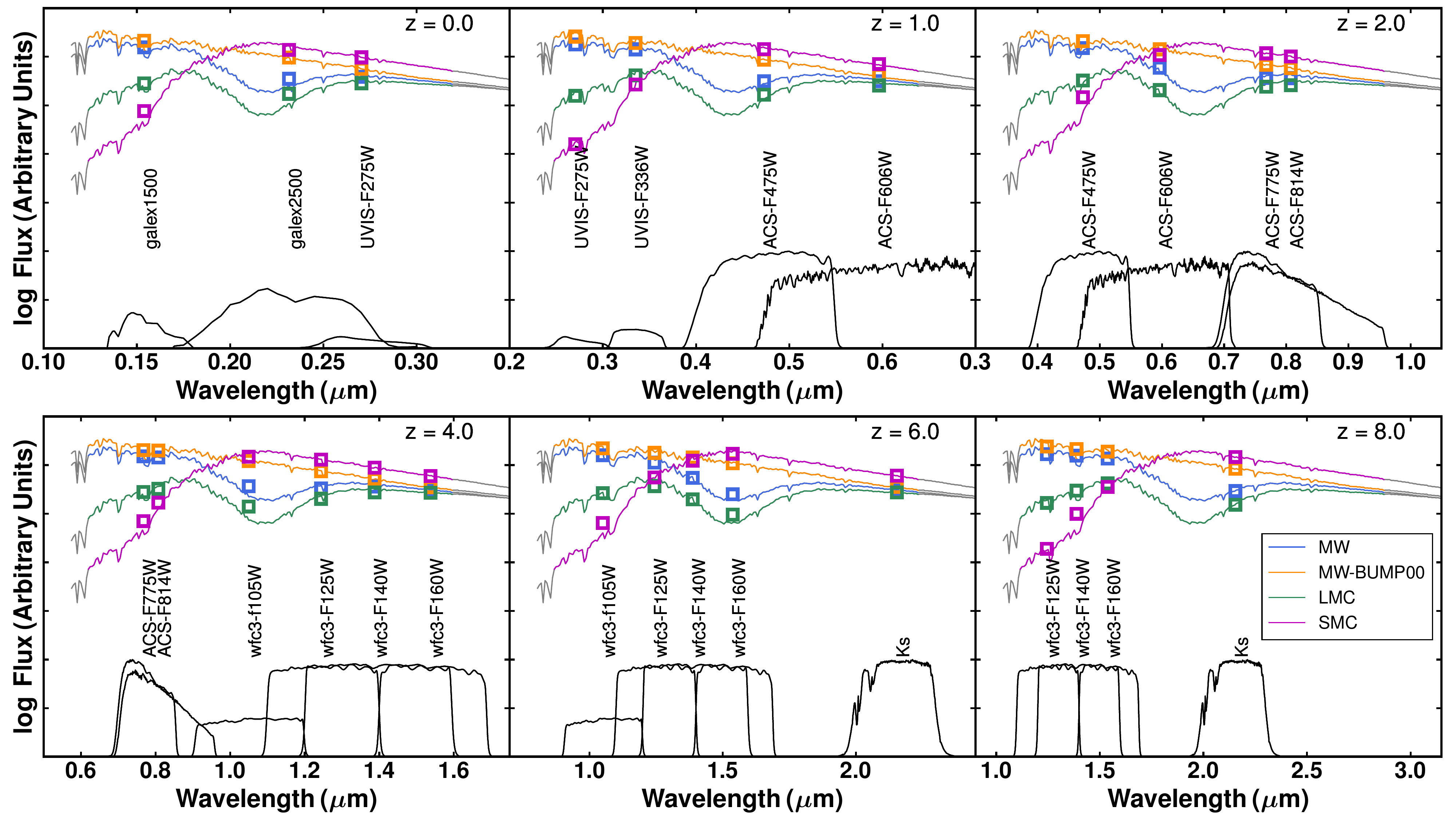}
\caption{The coverage of the UV rest-frame wavelength regime by
  different filters for redshifts running from $z=0$ to $z=8$, when
  adopting the four different dust-types in this study. The
  photometric sampling 
  includes the HST filters that are part of the \citet{Reddy2017}
  study (3D-HST filters and UVIS-F275W and UVIS-F336W), on top of
  \Galex, and a
  Ks-band filter. The coloured
  lines mark the attenuated spectra (from rest-frame 1230 to 3200 \Angstrom)
  assuming a V-band optical depth of
  $<\tau_{\text V}>=1$, whereas the coloured open squares mark the
  photometric measurements. The relevant filters at each redshift and their
  transmission curves are plotted below the stellar spectra. The
  photometric datapoints are marked as coloured squares, where the
  colours correspond to the attenuated spectra for different dust
  attenuation curves. 
\label{fig:observed_spectra_UVIS}}
\end{figure*}

\begin{figure*}
\includegraphics[width = 0.6\hsize]{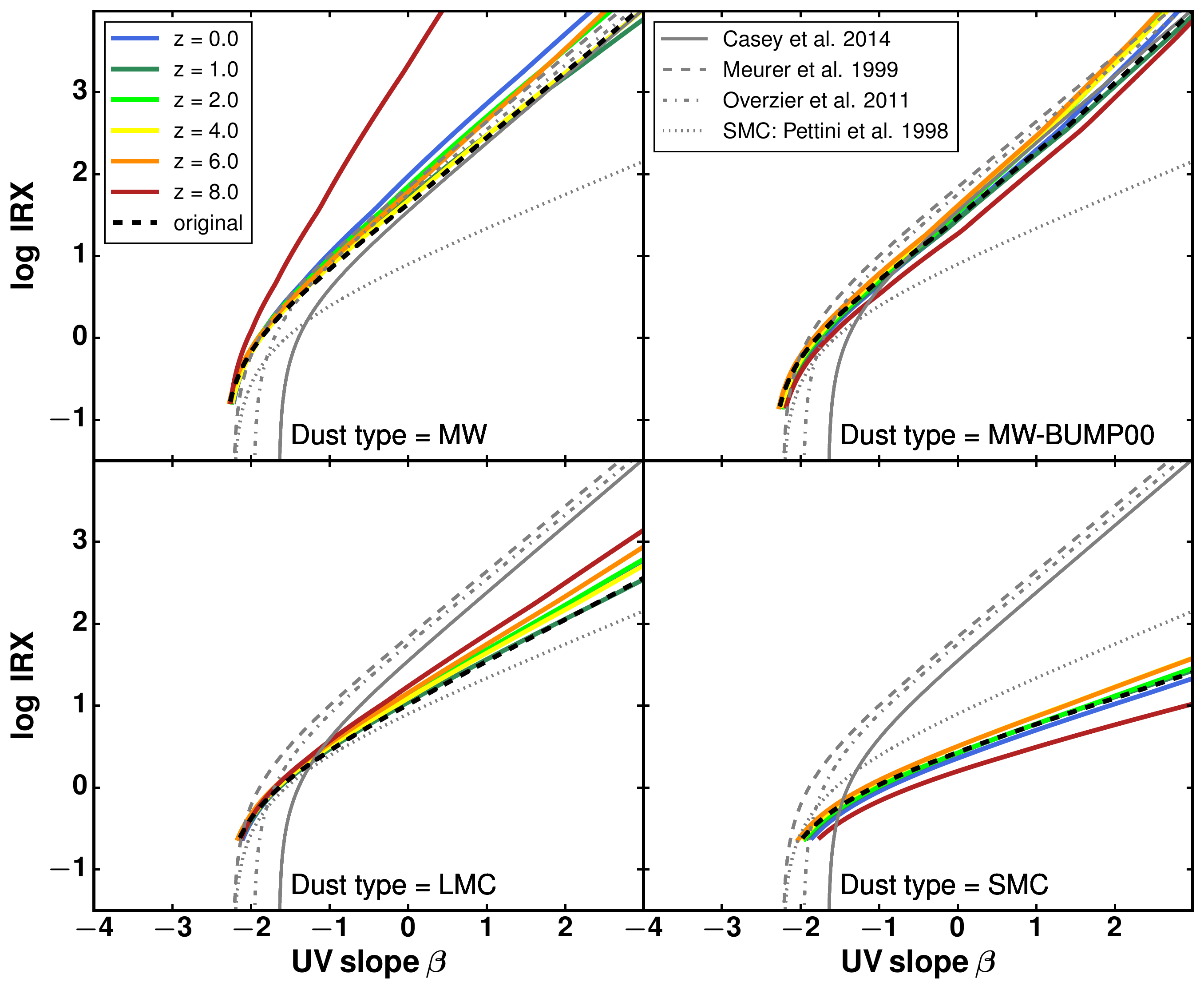}
\caption{The \IRXB relation of stellar emission from
  different redshifts based on a fit to photometric
  datapoints for different dust types. The
  photometric sampling 
  includes the HST filters that are part of the \citet{Reddy2017}
  study (3D-HST filters and UVIS-F275W and UVIS-F336W), on top of \Galex, and a
  Ks-band filter. Photometric datapoints were obtained by convolving the
  redshifted stellar spectrum with telescope filters corresponding to
  the rest-frame wavelength range $1230<\lambda<3200$.  The black dashed lines correspond to the original
  \IRXB relation derived from a direct fit to the attenuated
  spectrum. The 
  sampling of telescope filters of the UV wavelength range can
  completely alter the location of galaxies in the \IRXB plane for
  dust types with a steep UV-slope (SMC dust) or UV absorption
  features (2175 \Angstrom bump; MW and LMC dust). The better sampling
  of the UV part of the spectrum at e.g., $z=1$ improves the reconstruction
  of the \IRXB relation.
\label{fig:observed_beta_IRX_UVIS}}
\end{figure*}

Figure \ref{fig:observed_spectra_UVIS} shows the coverage of the UV
rest-frame wavelength regime using \Galex and a
Ks-band filter, as well as the filters that are part of the 3D-HST
survey \citep{Skelton2014} and two additional HST UVIS filters
\citep[UVIS-F275W and UVIS-F336W, as in][]{Reddy2017}. The addition of the
UVIS filters compared to the filter set used in Figure
\ref{fig:observed_spectra_3DHST} improves the sampling of the rest-frame
NUV spectrum at $z=0$ and the rest-frame FUV spectrum at $z=1$. This
is reflected in the derived \IRXB relation for this combination of
filters (Figure \ref{fig:observed_beta_IRX_UVIS}). Except for the
\IRXB relation at $z=8$ for a MW dust type, all `observed' \IRXB
relations based on a photometric sampling are nearly identical to the
\IRXB relation based on a direct fit to the rest-frame UV spectrum.

\begin{figure*}
\includegraphics[width = 0.8\hsize]{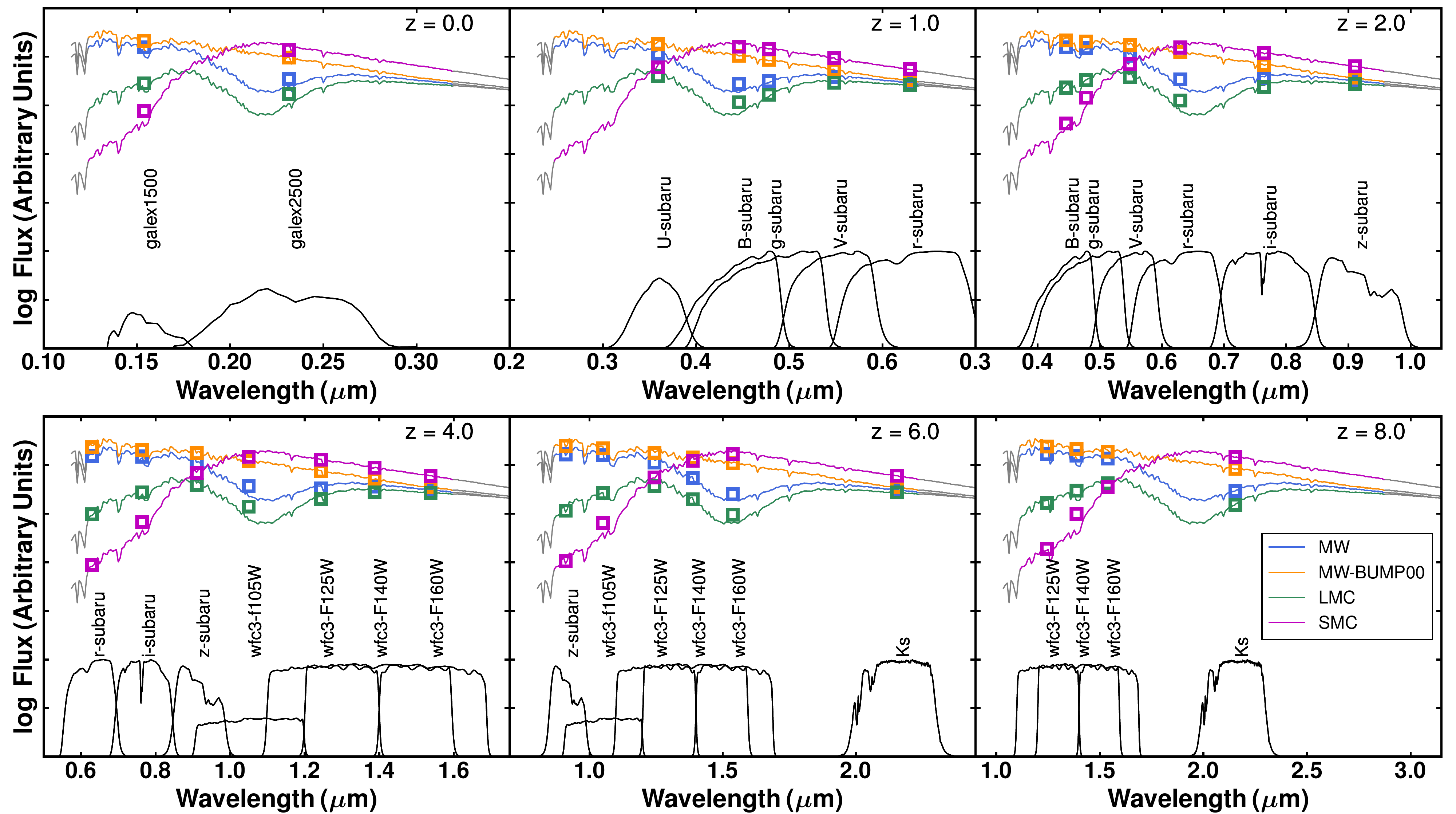}
\caption{The coverage of the UV rest-frame wavelength regime by
  different filters for redshifts running from $z=0$ to $z=8$, when
  adopting the four different dust-types in this study. The
  photometric sampling 
  includes the filters used in \citet[Subaru + HST WFC3]{Casey2014}, on top of \Galex,  and a
  Ks-band filter. The coloured
  lines mark the attenuated spectra (from rest-frame 1230 to 3200 \Angstrom)
  assuming a V-band optical depth of
  $<\tau_{\text V}>=1$, whereas the coloured open squares mark the
  photometric measurements. The relevant filters at each redshift and their
  transmission curves are plotted below the stellar spectra. The
  photometric datapoints are marked as coloured squares, where the
  colours correspond to the attenuated spectra for different dust
  attenuation curves. 
\label{fig:observed_spectra_Subaru}}
\end{figure*}

We repeat this exercise for a combination of filters from
\Galex, a Ks-band filter, HST WFC3 filters, and
Subaru filters \citep[this combination is based on the work
by][]{Casey2014}. This combination is affected by contamination by the
2175 \Angstrom bump at $z=0$ and $z=8$ and poorly samples the rest-frame FUV
regime of the attenuated stellar spectra of objects at $z=1$ (Figure
\ref{fig:observed_spectra_Subaru}). These effects lead to a poor
reconstruction of the \IRXB relation at $z=0$ and $z=8$ for  MW and
LMC dust types, as well as a poor reconstruction of the \IRXB relation
at $z=1$ for a SMC dust type.

To aid observers, Table \ref{tab:filter_combs} lists suggested filter
combinations as a function of redshift that reliably reconstruct
the UV slope $\beta$. To this aim, good photometric coverage of the FUV and the
NUV emission is necessary.

\begin{figure*}
\includegraphics[width = 0.6\hsize]{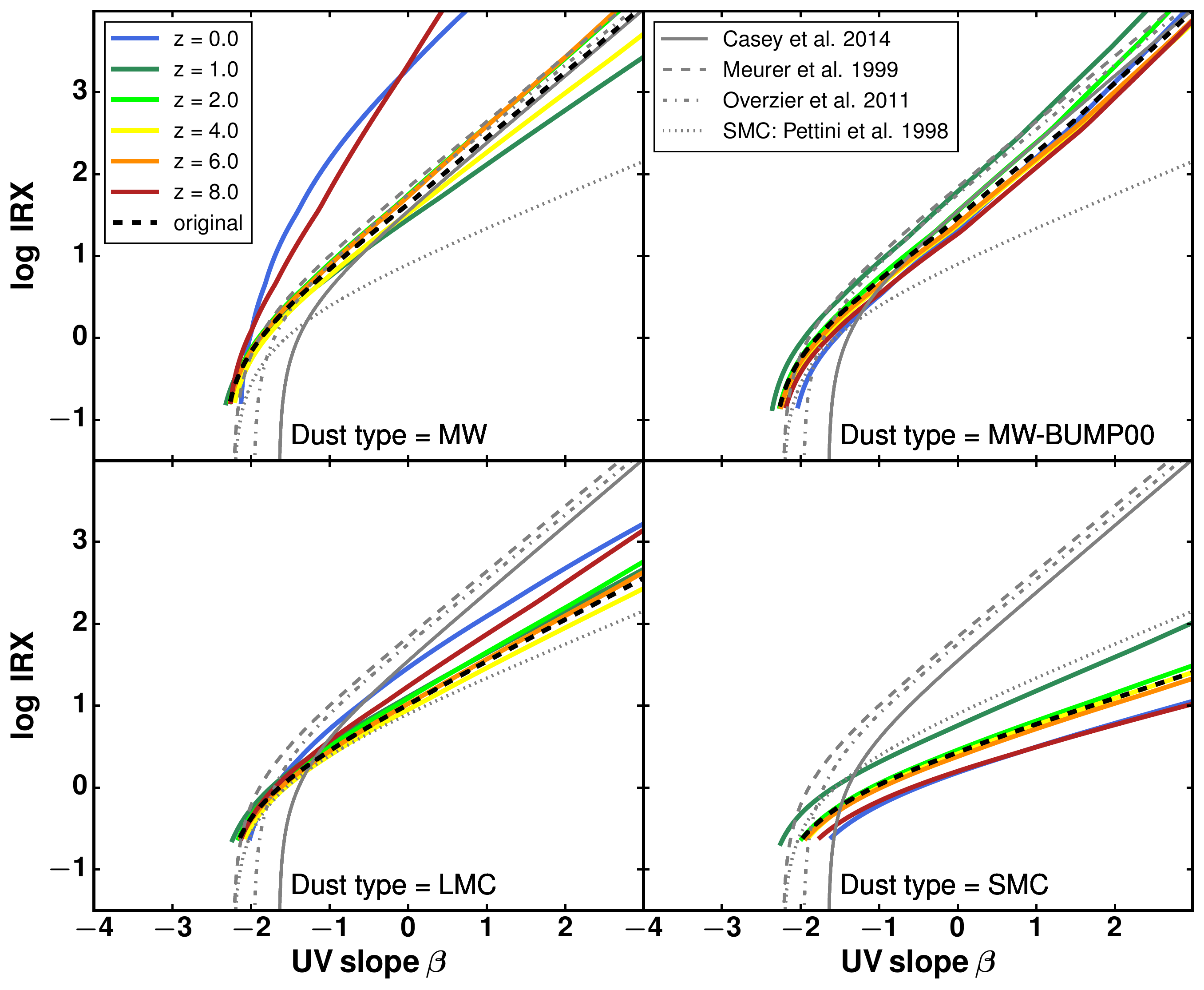}
\caption{The \IRXB relation of stellar emission from
  different redshifts based on a fit to photometric
  datapoints for different dust types. The
  photometric sampling 
  includes the filters used in \citet[Subaru + HST WFC3]{Casey2014}, on top of \Galex, and a
  Ks-band filter. Photometric datapoints were obtained by convolving the
  redshifted stellar spectrum with telescope filters corresponding to
  the rest-frame wavelength range $1230<\lambda<3200$.  The black dashed lines correspond to the original
  \IRXB relation derived from a direct fit to the attenuated
  spectrum. The 
  sampling of telescope filters of the UV wavelength range can
  completely alter the location of galaxies in the \IRXB plane for
  dust types with a steep UV-slope (SMC dust) or UV absorption
  features (2175 \Angstrom bump; MW and LMC dust). The improved
  sampling of the UV part of the rest-frame spectrum at e.g, $z=1$
  improves the reconstruction of the \IRXB relation.
\label{fig:observed_beta_IRX_Subaru}}
\end{figure*}

\end{document}